\documentclass[%
 aip,
 amsmath,amssymb,
reprint, twocolumn 
]{revtex4-1}

\usepackage{graphicx}
\usepackage{dcolumn}
\usepackage{bm}

\usepackage[utf8]{inputenc}
\usepackage[T1]{fontenc}
\usepackage{mathptmx}
\usepackage{etoolbox}
\usepackage{listings} 
\usepackage{amssymb,amstext,amsfonts,amsmath}
\usepackage[english]{babel}
\usepackage[usenames,dvipsnames]{color}

\makeatletter
\def\@email#1#2{%
 \endgroup
 \patchcmd{\titleblock@produce}
  {\frontmatter@RRAPformat}
  {\frontmatter@RRAPformat{\produce@RRAP{*#1\href{mailto:#2}{#2}}}\frontmatter@RRAPformat}
  {}{}
}%
\makeatother
\begin{document}


\title[Limitations of the rate-distribution formalism]{Limitations of the rate-distribution formalism in describing luminescence quenching in the presence of diffusion}


\author{Jakub Jędrak}
\email[]{jjedrak@ichf.edu.pl}

\affiliation{Institute of Physical Chemistry, Polish Academy of Sciences, ul. Kasprzaka 44/52, 01-224 Warsaw, Poland}

\author{Gonzalo Angulo}

\affiliation{Institute of Physical Chemistry, Polish Academy of Sciences, ul. Kasprzaka 44/52, 01-224 Warsaw, Poland}

%

\date{\today}

\begin{abstract}

When encountering complex fluorescence decays that deviate from exponentiality, a very appealing and powerful approach is to use lifetime (or equivalent rate constant) distributions. These are related by Laplace transform to multi-exponential functions, stretched exponentials, Becquerel's law, and others. In the case of bimolecular quenching, time-independent probability distributions of the rate constants have occasionally been used. Here we show that this mathematical formalism has a clear physical interpretation only when the fluorophore and quencher molecules are immobile, as in the solid state. However, such an interpretation is no longer possible once we consider the motion of fluorophores with respect to quenchers. Therefore, for systems in which the relative motion of fluorophores and quenchers cannot be neglected, it is not appropriate to use the time-independent continuous reaction rate or decay time distributions to describe, fit, or rationalize experimental results.

\end{abstract}

\maketitle

\section{Introduction}

The temporal evolution of luminescence depends on many different physicochemical phenomena that occur not only locally but also in the environment of the chromophore. \cite{birks1970photophysics, valeur2013molecular} The simplest dynamic behavior follows an exponential law and can be understood as a transition between two electronic levels. Considering the fact that measurements are usually performed on ensembles, the obtained observations are averages. In such systems, the basic experimental quantity is the luminescence decay, $I(t)$, between states of the same spin multiplicity, i.e. the fluorescence decay. We focus here on the fluorescence decay after an initial, very short pulse of light. 

Frequently, a finite number of exponents can be used to fit the experimental luminescence decay data: 
\begin{equation}
\label{I_multiexponential}
I(t) = \sum_{j=1}^{w} h_j e^{-t/\tilde{\tau}_j} = \sum_{j=1}^{w} h_j e^{ - \tilde{v}_j t}.
\end{equation}
In the above formula, $\tilde{\tau}_j$-s are decay times, while the parameters $\tilde{v}_j = 1/\tilde{\tau}_j$ are usually called {rate constants}, although the term {decay rates} seems more appropriate; $h_j$-s are the corresponding weights.

On occasion, however, strong deviations from the exponential or simple multiexponential decay law are observed. Their origins are varied: solvation dynamics, heterogeneous distribution of molecules, intramolecular excited state reactions, or diffusion-assisted bimolecular reactions. When one of these decreases the amount of observed luminescence with respect to a reference situation, it is referred to as quenching.
In many of these cases, a better fit to the experimental data is obtained by using other mathematical functions instead of (\ref{I_multiexponential}).\cite{berberan2005mathematical1, berberan2005mathematical2, berberan2007luminescence, berberan2008luminescence} These can be, for example, the single Becquerel function (compressed hyperbola) \cite{berberan2005mathematical2}, a weighted sum of Becquerel functions, a Kohlrausch function (stretched exponential), the unification of Becquerel and Kohlrausch functions \cite{berberan2005mathematical1, potopnyk2024aggregation}, or even more exotic functions like Mittag-Leffler or Heaviside’s exponential functions.  \cite{berberan2007luminescence, berberan2008luminescence, souchon2009multichromophoric, menezes2013methods} Therefore, instead of the multiexponential fit (\ref{I_multiexponential}), its generalization to the case of continuous rate constants or decay times is sometimes used. \cite{lakowicz1987analysis, vinogradov1994phosphorescence,wlodarczyk2006new, souchon2009multichromophoric, paulo2010single,fogarty2011extraction, menezes2013methods} In this approach, (\ref{I_multiexponential}) is replaced by
\begin{equation}
\label{I_expressed_by_H_tilde}
I(t) = \int_{0}^{\infty} \tilde{H}(\tilde{v}) e^{- \tilde{v} t} d \tilde{v} = \int_{0}^{\infty} \tilde{f}(\tilde{\tau}) e^{-t/\tilde{\tau}} d \tilde{\tau}.
\end{equation}
Importantly, both $\tilde{H}(\tilde{v})$ and $\tilde{f}(\tilde{\tau})$ are assumed to be time-independent.

The class of decay laws described by (\ref{I_expressed_by_H_tilde}) is quite large. For a suitable choice of the weighting function $\tilde{H}(\tilde{v})$ one obtains not only (\ref{I_multiexponential}), but also the Becquerel and Kohlrausch functions mentioned above, as well as many other functions as special cases. \cite{berberan2005mathematical1, berberan2005mathematical2, berberan2007luminescence, berberan2008luminescence} If $I(t)$ satisfies certain conditions, $\tilde{H}(\tilde{v})$ is non-negative and can be treated as a probability distribution. For this reason, we will use the term "rate distribution formalism" (RDF) for the approach that utilizes the equation (\ref{I_expressed_by_H_tilde}) to fit and rationalize the experimental data. This formalism is discussed in detail in the section \ref{RDF}.

The rate distribution formalism is still widely used (Ref. \onlinecite{berberan2005mathematical1} has more than 650 citations) to describe luminescence decay in various chemical systems. These include "fluorophores incorporated in micelles, cyclodextrins, rigid solutions, sol–gel matrices, polymers, proteins, vesicles or membranes, biological tissues, fluorophores adsorbed on surfaces, or linked to surfaces, quenching of fluorophores in micellar solutions, energy transfer in assemblies of like or unlike fluorophores, etc.".\cite{berberan2005mathematical1} In particular, we are interested in applications of the rate-distribution formalism as a description of the fluorescence quenching.

It follows from (\ref{I_expressed_by_H_tilde}) that $\tilde{H}(\tilde{v})$ can be obtained from the luminescence decay $I(t)$ by computing the inverse Laplace transform. This can be numerically challenging (the problem is ill-conditioned) and requires very good quality experimental data. \cite{berberan2008luminescence} However, even if one succeeds in obtaining $\tilde{H}(\tilde{v})$, there remains the question of the physical interpretation of the rate constants $\tilde{v}$.

Here we show that if the positions of fluorophores and quenchers are fixed (time-independent), then $\tilde{H}(\tilde{v})$ can be unambiguously determined, provided that we know both the functional form of the microscopic transfer rates and the spatial distribution of quenchers around the fluorophore (i.e., the fluorophore "microenvironment").

However, the time-independent distribution $\tilde{H}(\tilde{v})$ has no clear physical interpretation in any system where the relative motion of fluorophores and quenchers cannot be neglected. This includes not only simple liquids, but also most soft matter systems. Admittedly, in such cases one can use the relation (\ref{I_expressed_by_H_tilde}) to calculate $\tilde{H}(\tilde{v})$ from the experimental data on $I(t)$. But the quantity obtained in this way will no longer be the true distribution of rate constants resulting from detailed knowledge of microscopic details of the system: the transfer rates and "microenvironments" of the excited fluorophores.

To show all this, we use a theoretical framework of Ref. \onlinecite{dorfman1992influence}, which is equivalent to the Differential Encounter Theory (DET). \cite{doktorov1975quantum, burshtein1992geminate, burshtein2004non} However, our conclusions are quite general. In particular, they do not depend on the strength of iteractions between quenchers.

From a technical point of view, the explanation of the limitations of the rate-distribution formalism of Refs. \onlinecite{berberan2005mathematical1, berberan2005mathematical2, berberan2007luminescence, berberan2008luminescence} is the following: If the positions of quencher and fluorophore are fixed, the survival probability for a given excited fluorophore surrounded by any number of quenchers decreases exponentially with time, which corresponds to the term $\exp(-\tilde{v} t)$ in (\ref{I_expressed_by_H_tilde}). Precisely for this reason, the observation time $t$ can serve as a Laplace variable conjugated to $\tilde{v}$, but this is only possible for time-independent $\tilde{H}(\tilde{v})$.

In the general case, i.e. in the presence of motion (diffusive or otherwise) of the quenchers or fluorophores, the time dependence of the survival probability is no longer exponential. The rate-distribution formalism in the form presented in Refs. \onlinecite{berberan2005mathematical1,berberan2005mathematical2, berberan2007luminescence, berberan2008luminescence} is not sufficient to describe such a situation. Therefore, here we extend the original formulation of this theoretical framework. Namely, we construct distributions of rate constants that correctly incorporate microscopic details of the physical system in question, yet can be used in the presence of diffusion. However, such distributions are generally time-dependent, and in contrast to the time-independent distributions $\tilde{H}(\tilde{v})$ of Refs. \onlinecite{berberan2005mathematical1, berberan2005mathematical2, berberan2007luminescence, berberan2008luminescence}, they cannot be determined from the experimental luminescence decay data.
 

Everything that was said about rate constants $\tilde{v}$ applies also to the decay times $\tilde{\tau}$. This is because there is a simple one-to-one mapping between these two distributions variables and their probability distributions. Our main conclusion also holds in the discrete case: $\tilde{H}(\tilde{v}) = \sum_{j=1}^{w} h_j \delta(\tilde{v} - \tilde{v}_j)$ and $\tilde{f}(\tilde{\tau}) = \sum_{j=1}^{w} h_j \delta(\tilde{\tau} - \tilde{\tau}_j)$, which, when used in (\ref{I_expressed_by_H_tilde}) results in the equation (\ref{I_multiexponential}). Therefore, even the parameters of the simple multiexponential fit do not have a clear interpretation as soon as the fluorophores or quenchers are mobile.
 
Nevertheless, in the literature one can find examples of the application of rate or decay time distributions (either continuous or discrete) to describe luminescence decay in the presence of quenching molecules in liquids or soft matter systems. We found a number of such examples\cite{vinogradov1994phosphorescence, morandeira2003fluorescence, morandeira2004fluorescence, kuzmin2008evolution, paulo2010single, kuzmin2014analysis}, although we did not perform a systematic literature search.

\section{ Kinetics processes }

We consider here a system with fluorophores (F) and quenchers (Q) in liquid, soft matter or solid state. At $t=0$ a fraction of the fluorophore molecules are excited by the short light pulse: $ \text{F} + h \nu  \longrightarrow  \text{F}^{\ast}$. The excited fluorophore $\text{F}^{\ast}$ can return to the ground state either by spontaneous emission and internal conversion  
\begin{eqnarray}
\label{reaction_2nd_spontaneous_emission}
 \text{F}^{\ast} & \xrightarrow{1/\tau_0} & \text{F},
\end{eqnarray}
where $\tau_0$ is a natural lifetime of the excitation, or by reacting with the quencher. The quenching reaction may involve the transfer of energy ($E$), protons ($p^{+}$) or electrons ($e^{-}$), or other processes,  
\begin{eqnarray}
\label{reaction_3rd_quenching}
\text{F}^{\ast}  +  \text{Q}  & \xrightarrow{k_f(\mathbf{X})}  & \text{Products}.
\end{eqnarray}
In the above, $k_f(\mathbf{X})$ is the intrinsic rate of the elementary forward reaction responsible for the quenching. $\mathbf{X}$ is the vector of variables that are required to describe the position and orientation of a quencher relative to a fluorophore.

We assume that the quenching reaction (\ref{reaction_3rd_quenching}) is irreversible - there is no backward reaction leading to the fluorophore molecule in the excited state, $\text{F}^{\ast}$. However, there could be recombination leading to the neutral fluorophore in the ground state (geminate ion recombination in the case of electron transfer) or energy dissipation.  Nevertheless, the distribution of the rate constants $\tilde{H}(\tilde{v})$ in (\ref{I_expressed_by_H_tilde}) depends on the forward transfer (\ref{reaction_3rd_quenching}), but is not affected by the recombination process.

Let $N(t)$ be the  number of excited fluorophore molecules at time $t$. The basic experimental quantity of interest here is 
\begin{equation}
\label{definition_of_I}
\frac{N(t)}{N(0)} = I(t) =  \exp\left[-\frac{t}{\tau_0} -  c  \int_0^t k(t^{\prime}) d t^{\prime} \right].
\end{equation}
$I(t)$ (\ref{definition_of_I}) is called the 'decay law'\cite{berberan2008luminescence} or the 'fluorescence $\delta$-response function of a fluorescent sample'. \cite{boens2014identifiability} The last equality in (\ref{definition_of_I}) defines the time-dependent quenching rate $k(t)$; $c$ is the bulk quencher concentration. We assume here that $N(t)$ is large enough to be treated as a continuous variable. It is also assumed that $N(0) \ll Vc$, where $V$ is the total volume of our system.

$I(t)$ defined by (\ref{definition_of_I}) is the solution of the following equation:
\begin{equation}
\label{t_evolution_of_I}
\frac{d I(t)}{dt} =  -\frac{I(t)}{\tau_0} -  c  k(t) I(t),
\end{equation}
with the initial condition $I(0) = 1$. The time-dependent quenching rate $k(t)$ can be treated as a phenomelogical quantity defined by (\ref{definition_of_I}) or (\ref{t_evolution_of_I}), but it can also be computed using various theoretical approaches.\cite{dorfman1992influence, swallen1996solvent, doktorov1975quantum, burshtein1992geminate, burshtein2004non}

\section{Rate distribution formalism \label{RDF}}

In the introduction we briefly described the rate-distribution formalism (RDF).  \cite{berberan2005mathematical1, berberan2005mathematical2, berberan2007luminescence, berberan2008luminescence} We will now analyze this mathematical framework in more detail. 

After decomposing each rate constant or decay time into the part describing the natural decay of the excited state (\ref{reaction_2nd_spontaneous_emission}) and the remaining part: 
\begin{equation}
\label{definition_of_v_no_tilde}
\tilde{v}_j = \frac{1}{\tau_0}  + {v}_j = \frac{1}{\tau_0}  + \frac{1}{{\tau}_j} = \frac{1}{\tilde{\tau}_j},
\end{equation}
the equation (\ref{I_multiexponential}) takes the form
\begin{equation}
\label{I_multiexponential_bis_tau_0}
I(t) =  e^{-t/\tau_0}  \sum_{j=1}^{w} h_j e^{-t/\tau_j} = e^{-t/\tau_0} \sum_{j=1}^{w} h_j e^{ - v_j t}.
\end{equation}
From the definition (\ref{definition_of_I}) it follows that $I(0) = 1$ and hence $\sum_{j=1}^{w} h_j = 1$. For simplicity in what follows we will restrict our attention to the rate constants $v$ and their distributions. The corresponding distributions for the decay times $\tau$ can easily be obtained if needed.

Using (\ref{definition_of_v_no_tilde}) we can  rewrite (\ref{I_expressed_by_H_tilde}) as the continuous counterpart of (\ref{I_multiexponential_bis_tau_0}),
\begin{equation}
\label{I_expressed_by_H}
I(t) = e^{-\frac{t}{\tau_0}} \int_{0}^{\infty} H(v) e^{-v t} d v = e^{-\frac{t}{\tau_0}} \mathcal{L}[ H(v)](t),
\end{equation}
where  $v = \tilde{v} - 1/\tau_0$. Therefore, we have
\begin{equation}
\label{H_in_terms_of_I}
H_{}(v) = \mathcal{L}^{-1}[ e^{t/\tau_0} I_{}(t)](v) = \mathcal{L}^{-1}[\tilde{I}(t)](v). 
\end{equation}
$\mathcal{L}[\ldots]$ and $\mathcal{L}^{-1}[\ldots]$ denote the Laplace transform and its inverse, respectively. In (\ref{H_in_terms_of_I}) we also defined 
\begin{equation}
\label{definition_of_I_tilde}
\tilde{I}(t) \equiv I(t) e^{t/\tau_0}.
\end{equation}
$H(v)$ (\ref{H_in_terms_of_I}) differs from $\tilde{H}(\tilde{v})$ appearing in Eq. (\ref{I_expressed_by_H_tilde}) only by shift of the argument: $\tilde{H}(\tilde{v}) =  H\left(\tilde{v}- 1/\tau_0 \right)$. Clearly, (\ref{I_expressed_by_H}) is the generalization of the multiexponential function (\ref{I_multiexponential_bis_tau_0}), the latter being obtained for $H(v) = \sum_{j=1}^{w} h_j \delta({v} - {v}_j)$.

$H(v)$ appearing in (\ref{I_expressed_by_H}) is the central quantity of the RDF. If $\tilde{I}(t)$ (\ref{definition_of_I_tilde}) is completely monotone, \cite{berberan2008luminescence} that is, if for any $k=0, 1, \ldots$ and any $t$ we have
\begin{equation}
(-1)^{k} \frac{d^k \tilde{I}(t)}{d t^k}   > 0,  
\label{CM_definition}
\end{equation}
then ${H}({v})$ is non-negative and normalized, so it can be interpreted as probability distribution function (analogous condition for $\tilde{H}(\tilde{v})$ is $(-1)^{k}  I^{(k)}(t) > 0$). This is the case for example, for $\tilde{I}(t)$ described by the Becquerel function mentioned in the Introduction 
\begin{equation}
\label{Becquerel_I}
\tilde{I}(t) =  \frac{1}{(1+ (1-\gamma) t/\tau_c)^{1/(1-\gamma)}}, ~~~ 0 < \gamma \leq 1,
\end{equation}
for which the  distribution of rate constants ${H}({v})$ is the shifted gamma distribution. For a Kohlrausch function,
\begin{equation}
\label{Kohlrausch_I}
\tilde{I}(t) = \exp \left[- \left( \frac{t}{\tau_b} \right)^{\beta} \right], ~~~ 0 < \beta \leq 1,
\end{equation}
${H}({v})$ cannot be expressed in terms of elementary functions if $\beta \neq 1/2$. \cite{berberan2008luminescence} For $\beta  = 1/2$ we get
\begin{equation}
\label{Kohlrausch_H}
{H}({v}) = \frac{\tau_b}{2 \sqrt{\pi}(\tau_b v)^{3/2}} \exp \left(-\frac{1}{4 \tau_b v}\right).
\end{equation}
In (\ref{Becquerel_I}), (\ref{Kohlrausch_I}) and (\ref{Kohlrausch_H}), $\tau_b$ and $\tau_c$ are paramters with the dimension of time.

The essential feature of the original formulation of the RDF\cite{berberan2005mathematical1, berberan2005mathematical2, berberan2007luminescence, berberan2008luminescence} is that the distribution of the rate constants ${H}({v})$ does not depend on time. This makes it possible to compute ${H}({v})$ as an inverse Laplace transform, as in the examples above. However, within the formalism of Refs. \onlinecite{berberan2005mathematical1, berberan2005mathematical2, berberan2007luminescence, berberan2008luminescence},  the following time-dependent generalization of ${H}({v})$ is   also introduced:
\begin{equation}
\label{J_definition}
J(v, t) = \frac{H(v) e^{-v t}}{\int_{0}^{\infty} H(v) e^{-v t} d v} = \frac{H(v) e^{-v t}}{\tilde{I}(t)}.
\end{equation}
${J}({v}, t)$ obeys the following time-evolution equation
\begin{equation}
\label{time_evolution_equation_for_J}
\frac{\partial J(v, t)}{\partial t} = [c k(t) - v] J(v, t)
\end{equation}
with the initial condition
\begin{equation}
\label{J_initial_condition}
{J}({v}, 0) = {H}({v}).
\end{equation}
Comparing (\ref{definition_of_I}) and (\ref{I_expressed_by_H}) we get expressions for the time-dependent quenching rate $k(t)$, appearing in Eqs. (\ref{definition_of_I}) and (\ref{t_evolution_of_I}):
\begin{equation}
\label{k_as_average_of_nu_wrt_J}
c k(t) = \frac{\int_{0}^{\infty} v H(v) e^{-v t} d v}{\int_{0}^{\infty} H(v) e^{-v t} d v} = \int_{0}^{\infty} v J(v, t) d v.
\end{equation}
Note that the time dependence of ${J}({v}, t)$ is of a rather special form, which takes into account only the change in the number of excited fluorophores due to the first-order process characterized by the rate constant $v$: $H(v) \to H(v) e^{-v t}$. The denominator in Eq. (\ref{J_definition}) then ensures the correct normalization of ${J}({v}, t)$.  Such a form of ${J}({v}, t)$ does not allow to properly describe situations where the diffusive motion of fluorophores or quenchers cannot be neglected. (A generalization of ${J}({v}, t)$ that has a correct physical interpretation even in the presence of the relative motion of the fluorophor or quencher molecules will be derived in section \ref{correct_t_dep_H_and_J}).

However, in the case of immobile fluorophores and quenchers, the rate distribution formalism presented above can be given a precise physical interpretation in terms of microscopic quantities, see section \ref{Results}. Furthermore, RDF has an alternative derivation and interpretation, based on a particular system of first-order reactions, see Appendix \ref{Appendix_General_Scheme}. This heuristic derivation does not go into microscopic details of the system, but is very simple, intuitive, and completely general - in addition to fluorescence quenching, it can describe many other physical, chemical, or biochemical processes.

\section{Theoretical framework \label{Theory}}

In this section we present the mathematical formalism of Refs. \onlinecite{dorfman1992influence, swallen1996solvent}, which is equivalent to the Differential Encounter Theory (DET).\cite{doktorov1975quantum, burshtein1992geminate, burshtein2004non} The former approach is based on the backward Smoluchowski equation, which governs the time evolution of the survival probability for the fluorophore-quencher pair, i.e. the probability that in a system of one fluorophore and one quencher no transfer process has occurred by time $t$. This probability can then be used to construct the survival probability for a single fluorophore and an arbitrary number of quenchers.

In section \ref{Results} we will use the theoretical framework of Refs. \onlinecite{dorfman1992influence, swallen1996solvent} to show that the rate constants $v$ have no clear physical interpretation when the fluorophores or quenchers are mobile, while they have such an interpretation when the position and orientation of the fluorophores and quenchers do not change with time. Therefore, we argue that the rate distribution formalism of section \ref{RDF} should not be used when the effects of diffusion cannot be neglected. Then, in section \ref{correct_t_dep_H_and_J}, we show how to construct rate distributions that are valid in the case of mobile quenchers or fluorophores.

\subsection{Single fluorophore-quencher pair \label{single_pair}}

First, consider a single fluorophore-quencher pair in a medium where relative motion of an quencher with respect to a fluorophore is possible (i.e., liquid phase). The variables needed to describe the initial position and orientation of a quencher with respect to a fluorophore form a vector $\mathbf{X}$. In general, the components of $\mathbf{X}$ are the Cartesian coordinates and all angles necessary to describe the mutual orientation of the quencher and fluorophore molecules.

If we ignore for a moment the spontaneous fluorescence decay (\ref{reaction_2nd_spontaneous_emission}), then for the single fluorophore-quencher pair the time evolution of the survival probability $S^{(e)}_{1}(t|\mathbf{X})$ of the fluorophore excited state is given by the following backward Smoluchowski equation \cite{dorfman1992influence}
\begin{equation}
\label{S_1_BGSE}
\frac{\partial S^{(e)}_{1}(t|\mathbf{X})}{\partial t} = -k_f(\mathbf{X}) S^{(e)}_{1}(t|\mathbf{X}) + \hat{L}^{+}(\mathbf{X}) S^{(e)}_{1}(t|\mathbf{X}). 
\end{equation}
In the above equation, $k_f(\mathbf{X})$ is the forward electron or energy transfer rate (see Eq. (\ref{reaction_3rd_quenching})), whereas $\hat{L}^{+}(\mathbf{X})$ is the adjoint Smoluchowski operator describing the diffusive motion of the quencher with respect to the fluorophore. \cite{dorfman1992influence} In the simplest case of effectively spherical molecules and spherical symmetry of the system, $\mathbf{X}$ reduces to $R_0$: an initial (i.e., at $t=0$) fluorophore-quencher separation. In such a case we have\cite{goun2006photoinduced}
\begin{equation}
\label{S_1_BGSE_L_plus_for_spherical_symmetry}
\hat{L}_r^{+}(R_0) = \frac{1}{R^2_0}e^{V_r(R_0)}\frac{\partial}{\partial R_0}\left( D_r(R_0) R^2_0 e^{-V_r(R_0)} \frac{\partial}{\partial R_0}\right), 
\end{equation}
where $V_r(R_0)$ is a total interaction potential between fluorophore and quencher divided by $k_B T$.\footnote{Compared to (\ref{S_1_BGSE_L_plus_for_spherical_symmetry}), a different form of $\hat{L}_r^{+}(R_0)$ is used in Ref. \onlinecite{swallen1996solvent}:
\begin{equation}
\label{S_1_BGSE_L_for_spherical_symmetry_alternative}
\hat{L}_r^{+}(R_0) = \frac{D}{R^2_0}\exp[-V_r(R_0)]\frac{\partial}{\partial R_0} R^2_0 \exp[V_r(R_0)]\frac{\partial}{\partial R_0}. 
\end{equation}
It can be argued that the difference between (\ref{S_1_BGSE_L_plus_for_spherical_symmetry}) and (\ref{S_1_BGSE_L_for_spherical_symmetry_alternative}) is insignificant, since it is only a matter of sign convention and a way of defining $V_r(R_0)$. But this is true as long as we consider only the Coulomb interaction between fluorophore and quencher (if both are charged) and neglect the potential of the mean force. If the latter is included and defined as in (\ref{potential_of_the_mean_force}), then (\ref{S_1_BGSE_L_for_spherical_symmetry_alternative}) is incorrect. This is the case of Ref. \onlinecite{swallen1996solvent}, see section II B. of that Reference.
} We assume that there is no Coulomb interaction between the quencher and the fluorophore in the excited state, but we still include the 'potential of the mean force', defined as\cite{swallen1996solvent}
\begin{equation}
\label{potential_of_the_mean_force}
V_r(R_0) =  -\ln[g_r(R_0)],
\end{equation}
where $g_r(R_0)$ is the fluorophore-quencher radial pair distribution function. 
Note that if $V_r(R_0) \neq 0$ then the adjoint Smoluchowski operator $\hat{L}_r^{+}(R_0)$ (\ref{S_1_BGSE_L_plus_for_spherical_symmetry}) is different from the operator $\hat{L}_r(R)$ appearing in the corresponding forward Smoluchowski equation.\cite{dorfman1992influence}

\subsubsection{Initial and boundary conditions. The concept of a "bubble" \label{I_and_B_bubble}}

The equation (\ref{S_1_BGSE}) must be supplemented with the initial condition 
\begin{equation}
\label{S_1_BGSE_initial}
S^{(e)}_{1}(0|\mathbf{X}) = 1
\end{equation}
and the boundary conditions. The latter will be discussed here only for the simplest, spherically symmetric case when $S^{(e)}_{1}(t|\mathbf{X}) = S^{(e)}_{1r}(t|R_0)$, $k_f(\mathbf{X}) = k_{fr}(R_0)$ and $\hat{L}^{+}(\mathbf{X}) = \hat{L}_r^{+}(R_0)$. First, the flux must vanish at $R_0 = R_m$ (reflective boundary condition), where $R_m$ is the sum of the fluorophore and quencher radii (fluorophore-quencher contact distance), i.e.
\begin{equation}
\label{S_1_BGSE_boundary_inner_SS}
\frac{\partial}{\partial R_0}{S}^{(e)}_{1r}(t|R_0)_{\mid R_0 = R_m} =0.
\end{equation}
To discuss the outer boundary condition, note that
\begin{equation}
\label{S_1_BGSE_boundary_outer_lim_of_k_and_g}
\lim_{R_0 \to \infty} {g_r}(R_0) = 1, ~~~~ \lim_{R_0 \to \infty} k_{fr}(R_0) = 0.
\end{equation}
For any value of the observation time $t$ we also have
\begin{equation}
\label{S_1_BGSE_boundary_outer_lim_of_S}
\lim_{R_0 \to \infty} {S}^{(e)}_{1r}(t|R_0) = 1.
\end{equation}
We can choose $R_g$, $R_k$, and $R_{D}$ such that, to a good approximation, we have
\begin{equation}
\label{S_1_BGSE_boundary_outer_approx_of_k_and_g}
{g_r}(R_g) \approx 1, ~~~~  k_{fr}(R_k) \approx 0, ~~~~ {S}^{(e)}_{1r}(t|R_{D}) \approx 1.
\end{equation}
For each fluorophore excited at $t=0$ we can now define a 'bubble', i.e. all points lying inside a sphere with radius $R_b$. Our system contains $N(0)$ bubbles, each with a single fluorophore molecule in the center. We assume that the fluorophore concentration  is so low that we can ignore the interpenetration of the bubbles. This condition can only be strictly met in the solid state, where both fluorophores and quenchers are immobile. In such case we take
\begin{equation}
\label{definition_of_R_b}
R_b = \max(R_m, R_k)   ~~~ \text{(solids)}.
\end{equation}
In the case of liquids, the lack of bubble interpenetration can be a good approximation, but now we have to define $R_b$ as 
\begin{equation}
\label{definition_of_R_b_liquids}
R_b = R_{D}  ~~~ \text{(liquids)}
\end{equation}
where usually $R_{D} \gg \max(R_m, R_k)$. For liquids, the average fluorophore-fluorophore distance should also be larger than $2 R_b$. We assume that our system is diluted enough to satisfy this condition for any value of the observation time $t$ of interest, although $R_D$ clearly depends on both $t$ and the diffusion coefficient $D$. In addition, it follows from (\ref{S_1_BGSE_boundary_outer_approx_of_k_and_g}) and (\ref{definition_of_R_b_liquids}) that the net diffusive flux of quenchers across the bubble boundary is approximately zero.

Note that any model we use should be constructed in such a way that once we have chosen the radius of the bubble according to the formula (\ref{definition_of_R_b}) or (\ref{definition_of_R_b_liquids}), further increases in its size should not affect the values of macroscopic, observable physical quantities such as a luminescence decay $I(t)$ or a quenching rate $k(t)$.

\subsection{Single fluorophore and many quenchers \label{n_pairs}}

Now we are ready to discuss the case of a "bubble" with a fluorophore molecule in the center ($R_0 = 0$) and $n$ quenchers (labeled by $i = 1, 2, \ldots, n$) with initial distances from the fluorophore $R_{0i}$ obeying $R_{0i} < R_b$ for all $i$. 

We assume first that transfer process (\ref{reaction_3rd_quenching}) between a fluorophore and a given quencher is not altered in any way by the presence of other quenchers ('transfer independence'). Our second assumption is 'coordinate independence': we assume that the quencher molecules do not interact with each other, they "feel" the presence of the fluorophore only through the potential of the mean force (radial pair distribution function), Eq. (\ref{potential_of_the_mean_force}). Therefore, the coordinates $\mathbf{X}_i$ of the quencher molecules are not correlated. We still ignore a natural decay(\ref{reaction_2nd_spontaneous_emission}). 

With these assumptions, instead of (\ref{S_1_BGSE}), we have
\begin{equation}
\label{S_n_BGSE_no_tau_0}
\frac{\partial {S}^{(e)}_{n}(t|\mathbf{X}^n)}{\partial t} =  \sum_{i=1}^{n} \left[ \hat{L}^{+}(\mathbf{X}_i) - k_f(\mathbf{X}_i) \right] {S}^{(e)}_{n}(t|\mathbf{X}^n).  
\end{equation}
In the above, $\mathbf{X}_i$ describes the initial position and orientation of the $i$-th quencher and $\mathbf{X}^n = (\mathbf{X}_1, \mathbf{X}_2, \dots, \mathbf{X}_n)$. The form of the Smoluchowski operator $\hat{L}^{+}(\mathbf{X}_i)$ for the $i$-th quencher is consistent with the assumption of non-interacting quenchers; in the general case it would depend on the positions and orientations of other quencher molecules. The solution of (\ref{S_n_BGSE_no_tau_0}) has the following form 
\begin{equation}
\label{S_n_as_a_product_of_S_1}
{S}^{(e)}_{n}(t|\mathbf{X}^n) = \prod_{i=1}^{n}  S^{(e)}_{1}(t|\mathbf{X}_i), 
\end{equation}
i.e. the survival probability for the system of one fluorophore and $n$ quenchers is just a product of the $n$ independent survival probabilities for each fluorophore-quencher pair. Finally, taking into account the spontaneous transition from the excited state to the ground state of the fluorophore with fluorescence lifetime $\tau_0$, we obtain the full survival probability
\begin{equation}
\label{S_tilde_n_definition}
\tilde{S}^{(e)}_{n}(t|\mathbf{X}^n, \tau_0) = e^{-\frac{t}{\tau_0}} {S}^{(e)}_{n}(t|\mathbf{X}^n). 
\end{equation}
$\tilde{S}^{(e)}_{n}=\tilde{S}^{(e)}_{n}(t|\mathbf{X}^n, \tau_0)$ (\ref{S_tilde_n_definition}) obeys the following equation
\begin{equation}
\label{S_tilde_n_BGSE_with_tau_0}
\frac{\partial \tilde{S}^{(e)}_{n}}{\partial t} =  -\frac{1}{\tau_0} \tilde{S}^{(e)}_{n} + \sum_{i=1}^{n} \left[\hat{L}^{+}(\mathbf{X}_i) - k_f(\mathbf{X}_i) \right] \tilde{S}^{(e)}_{n}.  
\end{equation}
If there is no diffusion, $\hat{L}^{+}(\mathbf{X}_i) = 0$ for all $i$, the equations (\ref{S_1_BGSE}) and thus both (\ref{S_n_BGSE_no_tau_0}) and (\ref{S_tilde_n_BGSE_with_tau_0}) can be solved easily. We get 
\begin{equation}
\label{S_1_BGSE_solution_no_diffusion}
S^{(e)}_{1}(t|\mathbf{X}) = e^{- k_f(\mathbf{X}) t}, ~~~ {S}^{(e)}_{n}(t|\mathbf{X}^n) = \prod_{i=1}^{n}  e^{- k_f(\mathbf{X}_i) t}, 
\end{equation}
and therefore
\begin{equation}
\label{S_tilde_n_solution_no_diffusion}
\tilde{S}^{(e)}_{n}(t|\mathbf{X}^n, \tau_0) = e^{-t/\tau_0} \prod_{i=1}^{n}  e^{- k_f(\mathbf{X}_i) t}. 
\end{equation}
Importantly, (\ref{S_tilde_n_solution_no_diffusion}) is valid not only for the independent, randomly distributed quencher molecules, but also in a situation where the positions of the quenchers are arbitrarily correlated.

Now we need to relate the survival probabilities discussed above to the fluorescence  decay law $I(t)$. To find the relation in question, note that (\ref{definition_of_I}) can be rewritten as
\begin{equation}
\label{definition_of_I_with_I_n}
I(t) = \frac{N(t)}{N(0)} =  \frac{\sum_n N_n(t)}{N(0)} = \sum_n \frac{N_n(0)}{N(0)}  \frac{N_n(t)}{N_n(0)}= \sum_n a_n I_n(t). 
\end{equation}
In the above, $N_n(t)$ denotes the number of excited fluorophore molecules surrounded by a 'bubble' with exactly $n$ quenchers (the bubble concept is defined in subsection \ref{I_and_B_bubble}). Now  
\begin{equation}
\label{definition_of_a_n}
a_n = \frac{N_n(0)}{N(0)} 
\end{equation}
is a fraction of such $n$-quencher bubbles in our system, i.e. the probability of finding a bubble with $n$ quenchers when randomly drawing excited florophore at $t=0$. We also have 
\begin{equation}
\label{definition_of_I_n}
I_n(t) = \frac{N_n(t)}{N_n(0)} = \int_{B^n}  \tilde{S}^{(e)}_{n}(t|\mathbf{X}^n, \tau_0) q_n(\mathbf{X}^n) d \mathbf{X}^n, 
\end{equation}
where $q_n(\mathbf{X}^n)$ is the probability distribution of the initial coordinates of the $n$ molecules and $B^n = \underbrace{B \times B \times \ldots \times B}_n$. $I_n(t)$ (\ref{definition_of_I_n}) is the total survival probability for the fluorophore's excited state in the bubble with $n$ quenchers. Combining (\ref{definition_of_I_with_I_n}) and (\ref{definition_of_I_n}) we finally get 
\begin{equation}
\label{definition_of_I_with_S_tilde_n}
I(t) = \sum_n a_n I_n(t) = \sum_n a_n \int_{B^n}  \tilde{S}^{(e)}_{n}(t|\mathbf{X}^n, \tau_0) q_n(\mathbf{X}^n) d \mathbf{X}^n. 
\end{equation}
The above equation provides a link between observable quantity: decay of luminescence and microscopic characteristics of the system as given by $a_n$, $  \tilde{S}^{(e)}_{n}(t|\mathbf{X}^n, \tau_0)$ and $ q_n(\mathbf{X}^n)$.

\subsection{Liquid phase or disordered solid with uncorrelated initial positions of quencher molecules \label{Uncorrelated_q}}

Before we move to our results (Section \ref{Results}) we first need to consider the special situation when the intial positions of fluorophores and quenchers are uncorrelated, as is the case of liquid phase or a completely disordered solid. For this case, we re-derive the results known from the literature \cite{dorfman1992influence, swallen1996solvent}, in particular the formulas linking the luminescence decay $I(t)$ and the time-dependent quenching rate $k(t)$ to the conditional survival probability $S^{(e)}_{1}(t|\mathbf{X})$ for a single fluorophore-quencher pair. However, unlike e.g. Refs. \onlinecite{dorfman1992influence,swallen1996solvent},  we do not pass to the limit of infinite volume of the bubble. Instead, we explicitly use the Poisson distribution to model  the probability of having $n$ of quenchers in each bubble, see equation (\ref{Poisson_distribution}).

We assume that the quenchers interact with a fluorophore, but do not "feel" each other: their interactions are neglected.  Furthermore, although we consider the fluorophore-quencher excluded volume (through the presence of the pair distribution function $g(\mathbf{X})$), we neglect the quencher-quencher excluded volume effect. As a result, the quencher positions are assumed to be uncorrelated, and we have
\begin{equation}
\label{q_n_product_form}
q_n(\mathbf{X}^n) = \prod_{i=1}^{n} q_1 (\mathbf{X}_i). 
\end{equation}
If the positions of the quenchers are independent (as for the non-interacting molecules in solution), then the probability $a_n$ of finding $n$ quenchers in any bubble is given by a Poisson distribution \cite{van2007stochastic},
\begin{equation}
\label{Poisson_distribution}
a_n = \frac{\alpha^n}{n!}e^{-\alpha},
\end{equation}
where $\alpha$ is the average number of quenchers in a bubble: $\alpha = \sum_n n a_n$. Combining (\ref{S_n_as_a_product_of_S_1}), (\ref{S_tilde_n_definition}), (\ref{definition_of_I_with_S_tilde_n}), (\ref{q_n_product_form}) and (\ref{Poisson_distribution}) we get
\begin{eqnarray}
\label{definition_of_I_with_S_tilde_n_product_Poisson}
I(t) &=& \sum_n a_n I_n(t) \nonumber \\   &=& e^{-\frac{t}{\tau_0}} \sum_n \frac{\alpha^n}{n!}e^{-\alpha} \prod_{i=1}^{n} \int_{B} S^{(e)}_{1}(t|\mathbf{X}_i)  q_1 (\mathbf{X}_i)  d\mathbf{X}_i  \nonumber \\   &=& e^{-\frac{t}{\tau_0}} e^{-\alpha} \sum_n \frac{1}{n!} \left( \alpha \int_{B} S^{(e)}_{1}(t|\mathbf{X})  q_1 (\mathbf{X}) d\mathbf{X} \right)^n  \nonumber \\  &=&  \exp\left\{-\frac{t}{\tau_0} + \alpha \left[\int_{B} S^{(e)}_{1}(t|\mathbf{X})  q_1 (\mathbf{X})  d\mathbf{X} -1 \right] \right\}. \nonumber \\    
\end{eqnarray}
Probability distribution $q_1 (\mathbf{X}_i)$ is proportional to the radial (fluorophore -  quencher) pair distribution function $g (\mathbf{X}_i)$: 
\begin{equation}
\label{definition_of_q_1_with_g}
q_1 (\mathbf{X}) = \frac{g(\mathbf{X})}{\Omega_g},
\end{equation}
where 
\begin{eqnarray}
\label{definitions_Omega_g}
\Omega_g = \int_{B} g(\mathbf{X})d\mathbf{X}. 
\end{eqnarray}
We also define $C_V = N_A/V$, where $N_A$ is the total number of quenchers in a system and $V$ is the total system volume. $\Omega$ is the volume of the bubble and $C(\mathbf{X})= c g(\mathbf{X})$ is the local concentration field. Without loss of generality, for simplicity we consider here the case of a spherically symmetric "microenvironment" of the fluorophore and spherically symmetric molecules, hence $C(\mathbf{X})=c_r(R_0) = c g_r(R_b)$. We then assume that $\Omega_g$ defined by (\ref{definitions_Omega_g}) can be used instead of the volume of the bubble: $\Omega_g \approx \Omega$ and that $ C_V \approx c_r(R_b) \approx c$ (because $g_r(R_b) \approx 1$). In this case, $\alpha$ - the average number of quencher molecules in the bubble - can be replaced by $c \Omega_g$, 
\begin{equation}
\label{c_alpha_and_Omega_g}
\alpha = c \Omega_g. 
\end{equation}
Thus, from (\ref{definition_of_I_with_S_tilde_n_product_Poisson}) and (\ref{c_alpha_and_Omega_g}) we finally get
\begin{eqnarray}
\label{definition_of_I_with_S_tilde_n_product_Poisson_with_g}
I(t) &=& \exp\left[-\frac{t}{\tau_0} + c \int_{B} \left(S^{(e)}_{1}(t|\mathbf{X})-1 \right) g(\mathbf{X})  d\mathbf{X} \right]. 
\end{eqnarray}
Note that in (\ref{definition_of_I_with_S_tilde_n_product_Poisson}) the unphysical dependence of $\alpha $ and $q_1 (\mathbf{X})$ on $\Omega_g$ (or on the bubble radius $R_b$) cancels out. Therefore (\ref{definition_of_I_with_S_tilde_n_product_Poisson_with_g}) is expressed only by quantities like $c$, $S^{(e)}_{1}(t|\mathbf{X})$ or $g(\mathbf{X})$ which do not depend on the choice of the bubble boundary once the condition (\ref{definition_of_R_b}) or (\ref{definition_of_R_b_liquids}) is satisfied.

Comparing (\ref{definition_of_I_with_S_tilde_n_product_Poisson_with_g}) with Eq. (\ref{definition_of_I}) we obtain the following expression for the time-dependent quenching rate $k(t)$,
\begin{equation}
\label{definition_of_k_t}
k(t) = - \int_{B} \frac{\partial S^{(e)}_{1}(t|\mathbf{X})}{\partial t} g(\mathbf{X}) d\mathbf{X} =  \int_{B} k_f(\mathbf{X})  S^{(e)}_{1}(t|\mathbf{X}) g(\mathbf{X}). 
\end{equation}%

\section{Results \label{Results}}

In this section we show that when quenchers and fluorophores are immobile, the distribution of the rate constants $H(v)$ can be expressed in a simple way using the probabilities $a_n$, the probability distributions of the initial coordinates of the quenchers, $q_n(\mathbf{X}^n)$, and the forward transfer rate $k_f(\mathbf{X})$. However, if we consider the diffusion of either quencher or fluorophore molecules, then $H(v)$ determined by $a_n$, $q_n(\mathbf{X}^n)$ and $k_f(\mathbf{X})$ no longer agrees with the distribution of rate constants obtained from the observed luminescence decay $I(t)$. This is the central result of the present work.

\subsection{Dependence of the distribution of the rate constants $H(v)$ on the microenvironments of the fluorophores \label{Correlated_q}}

How to define $v$ and $H(v)$ using the transfer rates $k_f(\mathbf{X}_i)$ and the knowledge of the microscopic details of each fluorophore's molecular environment ('microenvironment'), defined by the distributions $a_n$ (\ref{definition_of_a_n}) and $q_n(\mathbf{X}^n)$? First, consider only the bubbles containing exactly $n$ quenchers. Then the rate constant $v$ should be defined as
\begin{equation}
\label{nu_definition_n_bubble}
v = \sum_{i=1}^{n}  k_f(\mathbf{X}_i).
\end{equation}
This is the only reasonable - and indeed inescapable - relation between the microscopic transfer rates $k_f(\mathbf{X}_i)$ and the observable decay rates $v$.

To define the distribution of the rate constants $H(v)$, we must first define the contribution to $H(v)$ of the fluorophores surrounded by exactly $n$ quenchers. This quantity is denoted here by $H_n(v)$, its natural definition is
\begin{equation}
\label{definition_of_H_n}
H_n(v) = \int_{B^n} \delta\left(v -\sum_{i=1}^{n}  k_f(\mathbf{X}_i)\right)   q_n(\mathbf{X}^n) d \mathbf{X}^n.
\end{equation}
In the above, $\delta(x)$ is the Dirac delta distribution. The equation (\ref{definition_of_H_n}) is the standard formula for changing the variables of the probability distribution \cite{van2007stochastic}, here from $\mathbf{X}^n = (\mathbf{X}_1, \mathbf{X}_2, \dots, \mathbf{X}_n)$ to $v$. Note again that $q_n(\mathbf{X}^n)$ does not have to be in the product form (\ref{q_n_product_form}), because in general case the initial coordinates of the quenchers do not have to be independent. We assume that the order of integration with respect to $\mathbf{X}^n$ and $v$ in (\ref{definition_of_H_n}) can be interchanged, and therefore for every $n$ the distribution $H_n(v)$ (\ref{definition_of_H_n}) is properly normalized:  $\int_{0}^{\infty}H_n(v) dv = 1$.

If there were only bubbles with exactly $n$ quenchers in a system (as might be the case in a simple crystalline solid), we would have $H_n(v) = H(v)$. In the general situation we have a contribution to $H(v)$ from bubbles (fluorophores) with different numbers $n$ of quenchers, and we get
\begin{equation}
\label{definition_of_H_with_H_n}
H(v) = \sum_n a_n H_n(v), 
\end{equation}
where $a_n$-s are the probabilities of finding the excited fluorophore surrounded by $n$ quenchers (i.e. the $n$ quencher bubble), as defined by (\ref{definition_of_a_n}). Of course, in the general case, $a_n$ does not have to be given by the Poisson distribution (\ref{Poisson_distribution}).

The formulas (\ref{definition_of_H_n}) and (\ref{definition_of_H_with_H_n}) express the dependence of $H(v)$ on the microenvironment of a fluorophore, but different combinations of $a_n$, $k_f(\mathbf{X})$ and $q_n(\mathbf{X}^n)$ can lead to the same $H(v)$.

Note that $H(v)$ defined by (\ref{definition_of_H_n}) and (\ref{definition_of_H_with_H_n}) is completely determined by the values of the physical quantities at $t=0$, knowledge of the values of these quantities at any $t>0$ is not required. This is consistent with the original formulation of the rate distribution formalism, \cite{berberan2005mathematical1, berberan2005mathematical2, berberan2007luminescence, berberan2008luminescence} where $H(v)$ is the initial condition for the time-dependent distribution $J({v}, t)$: ${J}({v}, 0) = {H}({v})$, see Eq. (\ref{J_initial_condition}).

\subsubsection{$H(v)$ has clear microscopic interpretation only when fluorophore and quencher are immobile \label{interpretation_clear_only_for_immobile}}

The distribution of the rate constants $H(v)$ defined by (\ref{definition_of_H_n}) and (\ref{definition_of_H_with_H_n}) leads to the correct formula for $I(t)$ as given by (\ref{definition_of_I_with_S_tilde_n}) only if the positions and orientations of the quencher or fluorophore molecules are fixed and do not change with time. Indeed, in such a case, using (\ref{S_1_BGSE_solution_no_diffusion}), (\ref{S_tilde_n_solution_no_diffusion}), (\ref{definition_of_I_with_S_tilde_n}), (\ref{definition_of_H_n}) and (\ref{definition_of_H_with_H_n}), we obtain
\begin{widetext}
\begin{eqnarray}
\label{definition_of_I_with_H_n}
e^{-\frac{t}{\tau_0}} \int_{0}^{\infty}  e^{-v t} H(v) d v   &=&  e^{-\frac{t}{\tau_0}}\sum_n a_n  \int_{0}^{\infty}  e^{-v t}  H_n(v) d v =  e^{-\frac{t}{\tau_0}} \sum_n a_n  \int_{0}^{\infty}   \int_{B^n}   \delta\left(v -\sum_{i=1}^{n}  k_f(\mathbf{X}_i)\right)  e^{-v t}   q_n(\mathbf{X}^n) d \mathbf{X}^n d v \nonumber \\
&=& e^{-\frac{t}{\tau_0}}  \sum_n a_n \int_{B^n} e^{ - t \sum_{i=1}^{n}  k_f(\mathbf{X}_i) } q_n(\mathbf{X}^n) d \mathbf{X}^n = e^{-\frac{t}{\tau_0}}  \sum_n a_n \int_{B^n} {S}^{(e)}_{n}(t|\mathbf{X}^n) q_n(\mathbf{X}^n) d \mathbf{X}^n \nonumber \\
&=& \sum_n a_n \int_{B^n}  \tilde{S}^{(e)}_{n}(t|\mathbf{X}^n, \tau_0) q_n(\mathbf{X}^n) d \mathbf{X}^n = \sum_n a_n I_n(t) = I(t).
\end{eqnarray}
\end{widetext}
For immobile quenchers or fluorophores we can replace $\prod_{i=1}^{n} e^{- k_f(\mathbf{X}_i) t} = e^{ - t \sum_{i=1}^{n}  k_f(\mathbf{X}_i) }$ by ${S}^{(e)}_{n}(t|\mathbf{X}^n)$, so the equation (\ref{I_expressed_by_H}) is indeed satisfied. Note that $q_n(\mathbf{X}^n)$ in (\ref{definition_of_I_with_H_n}) does not have to be of the product form (\ref{q_n_product_form}), because in the absence of diffusion we do not have to assume that the positions of the quenchers are uncorrelated to be able to use (\ref{S_tilde_n_solution_no_diffusion}).

But if the positions of fluorophore and quencher are not fixed, the solution of (\ref{S_1_BGSE}) is no longer of the simple exponential form (\ref{S_1_BGSE_solution_no_diffusion}): $S^{(e)}_{1}(t|\mathbf{X}) = e^{- k_f(\mathbf{X}) t}$ (\ref{S_1_BGSE_solution_no_diffusion}) does not obey (\ref{S_1_BGSE}) with $\hat{L}^{+}(\mathbf{X}) \neq 0$. So if $\hat{L}^{+}(\mathbf{X}) \neq 0$ then ${S}^{(e)}_{n}(t|\mathbf{X}^n) \neq \prod_{i=1}^{n} e^{- k_f(\mathbf{X}_i) t}$ and therefore $\tilde{S}^{(e)}_{n}(t|\mathbf{X}^n, \tau_0) \neq e^{-\frac{t}{\tau_0}} \prod_{i=1}^{n} e^{- k_f(\mathbf{X}_i) t}$. As a consequence, once diffusion cannot be neglected, the fundamental equation (\ref{I_expressed_by_H}) of the rate distribution formalism can no longer hold.

The last statement is true for a rather general form of the (adjoint) Smoluchowski operator $\hat{L}^{+}(\mathbf{X})$, i.e. in situations when the molecules can no longer be assumed to be effectively spherical or when rotational diffusion has to be taken into account. It also holds for the more general situation where quencher molecules interact with each other and the equation (\ref{S_n_BGSE_no_tau_0}) is replaced by  

\begin{equation}
\label{S_n_BGSE_no_tau_0_interacting_quenchers}
\frac{\partial {S}^{(e)}_{n}(t|\mathbf{X}^n)}{\partial t} =  \sum_{i=1}^{n} \left[ \hat{L}_i^{+}(\mathbf{X}^n) - k_f(\mathbf{X}_i) \right] {S}^{(e)}_{n}(t|\mathbf{X}^n).  
\end{equation}
In this case, the diffusion operator for the $i$-th quencher depends on the positions and orientations of the remaining quenchers (we recover the equation (\ref{S_n_BGSE_no_tau_0}) for non-interacting quenchers if we put $\hat{L}_i^{+}(\mathbf{X}^n) = \hat{L}^{+}(\mathbf{X}_i)$). As a consequence, ${S}^{(e)}_{n}(t|\mathbf{X}^n)$, which is the solution of (\ref{S_n_BGSE_no_tau_0_interacting_quenchers}), usually does not have the product form (\ref{S_n_as_a_product_of_S_1}). It is even more so clear that the solution of the backward Smoluchowski equation (\ref{S_n_BGSE_no_tau_0_interacting_quenchers}) in general does not have simple exponential form (\ref{S_1_BGSE_solution_no_diffusion}). ${S}^{(e)}_{n}(t|\mathbf{X}^n) = \prod_{i=1}^{n} e^{- k_f(\mathbf{X}_i) t}$ (\ref{S_1_BGSE_solution_no_diffusion}) would be a solution of (\ref{S_n_BGSE_no_tau_0_interacting_quenchers}) only if the following identity was true
\begin{equation}
\label{S_n_BGSE_no_tau_0_interacting_quenchers_no_k_f}
\sum_{i=1}^{n}  \hat{L}_i^{+}(\mathbf{X}^n) e^{ -\sum_{i=1}^{n}  k_f(\mathbf{X}_i) t}  = 0.  
\end{equation}
However, in general situation (e.g. for arbitrary value of time variable) this cannot be the case. Note that $\sum_{i=1}^{n}  \hat{L}_i^{+}(\mathbf{X}^n)$ does not depend on time, while ${S}^{(e)}_{n}(t|\mathbf{X}^n)$ does. Therefore, as we want to emphasize again, our conclusions remain valid for a rather general form of a diffusion operator $\hat{L}_i^{+}(\mathbf{X}^n)$. This means that the rate distribution formalism has no clear interpretation not only in the case of liquids, but also in the case of various soft matter systems (polymers, gels, liquid crystals), including biological systems.

\subsubsection{$D \to 0$ and $D \to \infty$ limits}

There is clearly one situation where the diffusive motion of the molecules is present but the rate distribution formalism is still applicable: the case of very slow diffusion ($D \approx 0$). In this limit ${S}^{(e)}_{n}(t|\mathbf{X}^n)$ (\ref{S_1_BGSE_solution_no_diffusion}) can be treated as an approximate solution of the Smoluchowski equation (\ref{S_n_BGSE_no_tau_0}) or (\ref{S_n_BGSE_no_tau_0_interacting_quenchers}).

More subtle is the case of infinitely fast diffusion ($D \to \infty$).\cite{agmon1983transient, rabinovich1990slow} More precisely, "infinitely fast" means that the diffusion is much faster than the transfer reaction (\ref{reaction_3rd_quenching}). In such a situation instead of (\ref{S_1_BGSE_solution_no_diffusion}) we have for all values of $\mathbf{X}^n$
\begin{equation}
\label{S_1_BGSE_solution_infinite_diffusion}
S^{(e)}(t|\mathbf{X}^n) = e^{- t \langle k_f(\mathbf{X}^n) \rangle_{\alpha}}. 
\end{equation}
In the above $\langle k_f(\mathbf{X}^n) \rangle_{\alpha}$ is the average not only over the equilibrium distribution of positions and orientations $\mathbf{X}^n$ of $n$ quenchers, but also over different values of $n$ -- the number of quenchers in a bubble. Explicitly, we have
\begin{eqnarray}
\label{average_k_f_infinite_diffusion_alpha}
\langle k_f(\mathbf{X}^n) \rangle_{\alpha} & = & \sum_n e^{-\alpha} \frac{\alpha^n}{n!} \int_{B^n}  \left[ \sum_{i=1}^{n}  k_f(\mathbf{X}_j) \right] \prod_{i=1}^{n} q_1 (\mathbf{X}_i)  d\mathbf{X}_i \nonumber \\ & = & \sum_n e^{-\alpha} \frac{\alpha^n}{n!} \int_{B} n   k_f(\mathbf{X})q_1 (\mathbf{X})  d\mathbf{X} = \alpha \langle k_f(\mathbf{X}) \rangle_{1}, \nonumber \\
\end{eqnarray}
where
\begin{eqnarray}
\label{average_k_f_infinite_diffusion_1}
\langle k_f(\mathbf{X}) \rangle_{1} & = & \int_{B}  k_f(\mathbf{X})q_1 (\mathbf{X})  d\mathbf{X}.
\end{eqnarray}
The equation (\ref{S_1_BGSE_solution_infinite_diffusion}) can be derived from the following intuitive, heuristic arguments. In the $D \to \infty$ limit, we  first average $k_f(\mathbf{X}^n)$ and only then we solve the time evolution equation for the survival probability, which now reads
\begin{equation}
\label{S_BGSE_infinite_diffusion}
\frac{\partial S^{(e)}(t|\mathbf{X}^n)}{\partial t} = -\langle k_f(\mathbf{X}^n) \rangle_{{\alpha}} S^{(e)}(t|\mathbf{X}^n). 
\end{equation}
In the present situation $n$ is not well defined, since we have to assume that the number of quenchers in each bubble varies rapidly (hence $S^{(e)}(t|\mathbf{X}^n)$ (\ref{S_1_BGSE_solution_infinite_diffusion}) has no $n$ lower index). But all bubbles are equivalent, and all fluorophores have (on average) exactly the same environment. Also, no matter what $n$ and $\mathbf{X}^n$ we start from, we have the same solution, because the memory of the initial conditions is lost immediately. Still, we assume that the quenchers move independently and that there is no correlation between their positions or orientations.

From what has been said above, it follows that in the present situation we have a simple mono-exponential luminescence decay,
\begin{equation}
\label{I_D_infity}
I(t) = \frac{N(t)}{N(0)} = \exp \left( - t \langle k_f(\mathbf{X}^n) \rangle_{\alpha}\right).
\end{equation}
Note that $\langle k_f(\mathbf{X}^n) \rangle_{\alpha} = c \int_{B} k_f(\mathbf{X}) g(\mathbf{X}) d\mathbf{X}$, so it does not depend on the choice of the bubble boundary, provided that the condition (\ref{definition_of_R_b}) or (\ref{definition_of_R_b_liquids}) is satisfied. 

But what about the applicability of the rate distribution formalism in the present case? Do rate constants and their distribution have a well-defined meaning also for the $D \to \infty$ limit? Since the survival probability (\ref{S_1_BGSE_solution_infinite_diffusion}) is not of the form (\ref{S_1_BGSE_solution_no_diffusion}), the equation (\ref{definition_of_I_with_H_n}) is, strictly spekaing not satisfied, despite the simple exponential form of (\ref{S_1_BGSE_solution_infinite_diffusion}). Still, one can argue that in the limit in question, instead of (\ref{nu_definition_n_bubble}), we should define the observable rate constants using the following formula  
\begin{equation}
\label{definition_v_D_infity}
v = \langle k_f(\mathbf{X}^n) \rangle_{\alpha},
\end{equation}
where $\langle k_f(\mathbf{X}^n) \rangle_{\alpha}$ is given by (\ref{average_k_f_infinite_diffusion_alpha}). If this is the case, then the rate distribution formalism is still applicable, and we get a (rather trivial) rate distribution corresponding to mono-exponential decay (\ref{I_D_infity}),
\begin{equation}
\label{definition_of_H_infinite_diffusion}
H(v) =  \delta\left(v - \langle k_f(\mathbf{X}^n) \rangle_{\alpha} \right).
\end{equation}
Note, however, that in the $D \to \infty$ limit the validity of the rate distribution formalisms depends entirely on the alternative definition (\ref{definition_v_D_infity}) of the macroscopic rate constant $v$.

\subsubsection{Deterministic quencher motion}

Our conclusions hold not only for diffusive motion, but also for deterministic motion of an quencher with respect to the fluorophore along a given trajectory $\mathbf{Z}(t)$. For $n$ quenchers we have $n$ such trajectories and the corresponding motions: $\mathbf{Z}_1(t), \mathbf{Z}_2(t), \dots, \mathbf{Z}_n(t)$. These can be solutions of any equations of motion, however complicated, with arbitrary interactions of quenchers, etc., and with the initial conditions $\mathbf{Z}_i(0) = \mathbf{X}_i$ for each $i = 1, 2, \ldots, n$. Now, instead of (\ref{S_n_BGSE_no_tau_0}), ${S}^{(e)}_{n}(t|\mathbf{X}^n)$ obeys the following equation
\begin{equation}
\label{S_n_BGSE_no_tau_0_det_motion}
\frac{d {S}^{(e)}_{n}(t|\mathbf{X}^n)}{d t} = - \sum_{i=1}^{n}   k_f \left[ \mathbf{Z}_{ i}(t) \right] {S}^{(e)}_{n}(t|\mathbf{X}^n).  
\end{equation}
Therefore we have
\begin{equation}
\label{S_1_BGSE_solution_no_diffusion_det_motion}
{S}^{(e)}_{n}(t|\mathbf{X}^n) = \prod_{i=1}^{n}  e^{- \int_0^t k_f \left[ \mathbf{Z}_{ i}(t^\prime) \right] dt^\prime}. 
\end{equation}
Note that the simple product form of ${S}^{(e)}_{n}(t|\mathbf{X}^n)$ (\ref{S_1_BGSE_solution_no_diffusion_det_motion}) relies on the fact that the electron or energy transfer from a fluorophore to a given quencher is in no way affected by the presence of other quenchers ('transfer independence' mentioned in subsection \ref{n_pairs}). However, we must assume that the quenching reaction (\ref{reaction_3rd_quenching}) is much faster than molecular motion. More precisely, we assume that the relative position and orientation of the fluorophore and the quencher will not change significantly during the time corresponding to the transfer process.

In principle one can imagine trajectories $\mathbf{Z}_1(t), \mathbf{Z}_2(t), \dots, \mathbf{Z}_n(t)$ which are chosen in such a way that $\sum_{i=1}^{n} k_f \left[ \mathbf{Z}_{i}(t) \right]$ remains constant. Then (\ref{S_1_BGSE_solution_no_diffusion_det_motion}) reduces to (\ref{S_1_BGSE_solution_no_diffusion}). But this is clearly not true in general, so ${S}^{(e)}_{n}(t|\mathbf{X}^n)$ (\ref{S_1_BGSE_solution_no_diffusion_det_motion}) is most likely not of simple exponential form (\ref{S_1_BGSE_solution_no_diffusion}). 

Thus, in the general case there is no direct connection between the microscopic parameters and the rate constants $v$, and hence the rate constants have no clear physical interpretation. Once again, we see that when fluorophore or quencher molecules become mobile, the rate distribution formalism of Refs. \onlinecite{berberan2005mathematical1, berberan2005mathematical2, berberan2007luminescence, berberan2008luminescence} is no longer applicable.

\subsubsection{Interim summary: difference between microscopic and observable quantities}

For clarity, in this Subsection we denote by $H_{\text{obs}}(v)$ the distribution of rate constants obtained from the experimental values of the luminescence decay $I(t) = I_{\text{obs}}(t) $ using Eq. (\ref{H_in_terms_of_I}). In contrast, $H_{\text{mic}}(v)$ denotes the 'microscopic' distribution of rate constants, which can be \textit{computed} using Eqs. (\ref{definition_of_H_n}) and (\ref{definition_of_H_with_H_n}), thus depending on our knowledge of the microscopic transfer rates $k_f(\mathbf{X})$ and the 'microenvironment' of each fluorophore given by $a_n$ and $q_n(\mathbf{X}^n)$.

The discussion of Subsection \ref{interpretation_clear_only_for_immobile} can be now rephrased and summarized as follows: with the help of equation (\ref{H_in_terms_of_I}) one can use luminescence decay obtained from the experiment ($I_{\text{obs}}(t)$) to determine the following distribution of rate constants, 
\begin{eqnarray}
\label{H_obs_in_terms_of_I_obs}
H_{\text{obs}}(v)  & = &  \mathcal{L}^{-1}[ e^{t/\tau_0} I_{\text{obs}}(t)]. 
\end{eqnarray}
We can also use (\ref{I_expressed_by_H}) to \textit{predict} the time dependence of the luminescence decay $I(t) = I_{\text{mic}}(t)$ by assuming a certain functional form of $H(v) = H_{\text{mic}}(v)$, i.e, by choosing a certain model for the transfer rates $k_f(\mathbf{X})$ and the distributions $a_n$ and $q_n(\mathbf{X}^n)$,
\begin{eqnarray}
\label{I_mic_obs_in_terms_of_H_mic}
I_{\text{mic}}(t)  & = &   e^{-\frac{t}{\tau_0}} \int_{0}^{\infty}  e^{-v t} H_{\text{mic}}(v) d v.
\end{eqnarray}
But if either the quencher or the fluorophore is mobile, we usually have 
\begin{eqnarray}
\label{H_obs_not_equal_to_H_mic}
H_{\text{mic}}(v) & \neq & H_{\text{obs}}(v),
\end{eqnarray}
hence $I_{\text{mic}}(t) \neq  I_{\text{obs}}(t)$. Note that $H_{{\text{mic}}}(v)$ is completely determined by $a_n$, $k_f(\mathbf{X})$ and $q_n(\mathbf{X}^n)$, which appear in the formulas (\ref{definition_of_H_n}) and (\ref{definition_of_H_with_H_n}) but does not depend on the details of the diffusion operator $\hat{L}^{+}(\mathbf{X}^n)$. In particular, $H_{{\text{mic}}}(v)$ does not depend on the diffusion coefficient. On the other hand, $H_{\text{obs}}(v)$ determined from the experimental data using (\ref{H_obs_in_terms_of_I_obs}) will generally depend on the diffusion coefficient $D$. If we change the viscosity of the solvent and thus the value of $D$, the time evolution of the system will also change. For different values of $D$ we get different experimental data on $I_{\text{obs}}(t)$, and this dependence of $I_{\text{obs}}(t)$ on $D$ is naturally carried over to $H_{\text{obs}}(v)$ by (\ref{H_obs_in_terms_of_I_obs}). Therefore, only in the case of immobile fluorophore and quencher molecules we can hope to get $H_{\text{mic}}(v) = H_{\text{obs}}(v)$, provided we model $a_n$, $k_f(\mathbf{X})$ and $q_n(\mathbf{X}^n)$ correctly.

\subsubsection{Our conclusions are also valid for the intramolecular transfer}

The conclusions reached in this paper also hold when the quencher and fluorophore are different fragments of a macromolecule, such as a protein. The relative motion of fluorophore and quencher is then related to internal motions within that macromolecule. Such a situation has been described and studied e.g. in Ref. \onlinecite{agmon1983transient}. There we have the following passage: 'As a concrete example consider an electron transfer reaction which depends exponentially on distance. Suppose that the electron fluorophore and electron quencher are imbedded inside a large protein. In such a macromolecule many bonds can rotate, bend, or vibrate, leading to large fluctuations of fluorophore-quencher distance. The random variations in this distance can be described by a diffusion equation. These variations are coupled to the electron transfer process because of the distance dependence of the rate constant.'. \cite{agmon1983transient}

Whether we describe this motion as diffusion or as deterministic motion does not change our main conclusion: the description of such a system within the rate-distribution formalism has no clear physical interpretation, just as in the situation where the fluorophore and quencher are separate molecules.

\subsection{How to construct rate distributions when fluorophore and quencher diffusion cannot be neglected? \label{correct_t_dep_H_and_J}}

Within the formalism of rate distributions, besides the distribution of the rate constant $H(v)$ (\ref{H_in_terms_of_I}), also its time-dependent generalization $J(v, t)$ (\ref{J_definition}) is defined.\cite{berberan2008luminescence} As in the case of $H(v)$, we will show here that $J(v, t)$ defined by (\ref{J_definition}), with its very special form of the time dependence has no clear physical interpretation if the fluorophores and quenchers are mobile. We will also show how to define the correct generalization of $J(v, t)$, which can be used even in the presence of mutual motion of fluorophore and quencher.

This is done in two steps. First, we define $\mathcal{H}(v, t)$ - the time-dependent generalization of $H(v)$. This $\mathcal{H}(v, t)$ reduces to $H(v)e^{-v t}$ in the case of immobile fluorophores and quenchers. Therefore, $\mathcal{H}(v, t)$ is not normalized to unity: $\int_0^{\infty} \mathcal{H}(v, t) dv < 1$ for $t>0$, so it is more appropriate to call it a pseudo probability distribution. Then we introduce the normalized counterpart of $\mathcal{H}(v, t)$: $\mathcal{J}(v, t)$, which is a direct generalization of $J(v, t)$ (\ref{J_definition}). For simplicity, we assume in this section that the quenchers do not interact, so their positions and orientations are independent.

\subsubsection{Preliminary definitions}

For our present purposes, it is convenient to use not only the survival probabilities $S^{(e)}_{1}(t|\mathbf{X})$ obeying the Backward Smoluchowski Equation (\ref{S_1_BGSE}), which we have used so far, but also the pair correlation function $n(\mathbf{X}, t)$ of the Differential Encounter Theory. \cite{doktorov1975quantum, burshtein1992geminate, burshtein2004non, angulo2020bimolecular} These two quantities are related by the following equation\footnote{In the original formulation of Differential Encounter Theory \cite{doktorov1975quantum} it was assumed that $n(\mathbf{X}, 0) = g(\mathbf{X}) =1$. However, DET can be easily generalized to include non-zero interaction potential between excited fluorophore and quenchers, and then we have $g(\mathbf{X}) \neq 1$, see for example Ref. \onlinecite{angulo2020bimolecular}.}
\begin{equation}
\label{S_and_n_connection}
S^{(e)}_{1}(t|\mathbf{X}) = \frac{n(\mathbf{X}, t)}{n(\mathbf{X}, 0)} = \frac{n(\mathbf{X}, t)}{g(\mathbf{X})},
\end{equation}
where $g(\mathbf{X})$ is the pair distribution function. The mapping between the Differential Encounter Theory (DET) and the formalism based on the Backward Smoluchowski Equation (BSE) allows one to work with the standard diffusion operator and not its adjoint form. For $g(\mathbf{X}) = 1$ we have $S^{(e)}_{1}(t|\mathbf{X}) = n(\mathbf{X}, t)$ and $\hat{L}^{+}(\mathbf{X}) = \hat{L}(\mathbf{X})$, so these two approaches lead to exactly the same equations.

To construct $\mathcal{H}(v, t)$ and $\mathcal{J}(v, t)$, we first define the following probability distributions
\begin{eqnarray}
\label{definition_of_q_1_or_p_of_t}
p(\mathbf{X}, t) &=&  \frac{ n(\mathbf{X}, t)}{\int_{B}   n(\mathbf{X}, 0)  d\mathbf{X}}    \equiv \frac{ n(\mathbf{X}, t)}{\Omega_g}    
\end{eqnarray}
and
\begin{eqnarray}
\label{definition_of_phi_of_t}
\phi(\mathbf{X}, t) &=&  \frac{ p(\mathbf{X}, t)}{\int_{B}   p(\mathbf{X}, t)  d\mathbf{X}}    \equiv \frac{ p(\mathbf{X}, t)}{Q(t)}.    
\end{eqnarray}
These two equations also define the quantities $\Omega_g$ (already defined by Eq. (\ref{definitions_Omega_g}) in section \ref{Uncorrelated_q})) and $Q(t)$. By invoking (\ref{S_and_n_connection}) and (\ref{definition_of_q_1_or_p_of_t}), we see that $Q(t)$ is the total survival probability for a single fluorophore-quencher pair, i.e., the probability that no quenching reaction involving this pair has occurred by time $t$,  \cite{agmon1983transient}
\begin{eqnarray}
\label{definition_of_Q_of_t}
Q(t) &=& \int_{B}   p(\mathbf{X}, t)  d\mathbf{X} =  \int_{B}  S^{(e)}_{1}(t|\mathbf{X}) q_1(\mathbf{X}) d\mathbf{X}.
\end{eqnarray}
This follows from the fact that
\begin{equation}
\label{q_1_is_p_at_t_0}
q_1(\mathbf{X}) = \Omega_g^{-1} n(\mathbf{X}, 0) = p(\mathbf{X}, 0) = \phi(\mathbf{X}, 0)
\end{equation}
is the probability distribution of the initial coordinates of the quencher, see Eq.  (\ref{definition_of_q_1_with_g}). Note that $\phi(\mathbf{X}, t)$ (\ref{definition_of_phi_of_t}) is normalized to unity, $\int_{B}   \phi(\mathbf{X}, t) d\mathbf{X} = 1$ while $p(\mathbf{X}, t)$ (\ref{definition_of_q_1_or_p_of_t}) is not: $\int_{B} p(\mathbf{X}, 0) d\mathbf{X} = 1$ but $\int_{B} p(\mathbf{X}, t) d\mathbf{X} \leq 1$ for $t>0$. Nevertheless, both $p(\mathbf{X}, t)$ and $\phi(\mathbf{X}, t)$ describe the distribution of quenchers around a fluorophore at time $t$.

\subsubsection{Definitions of $\mathcal{H}(v, t)$ and $\mathcal{J}(v, t)$}

Now we are ready to define the time-dependent counterpart of $H(v)$, which is also a generalization of $H(v)e^{-vt}$: 
\begin{equation}
\label{definition_of_H_t_with_H_n_t}
\mathcal{H}(v, t)= \sum_n \tilde{a}_n(t) \mathcal{H}_n(v, t). 
\end{equation}
In the above, $\mathcal{H}_n(v, t)$ is the time-dependent counterpart  of $H_n(v)$ (\ref{definition_of_H_n}) and is given by
\begin{equation}
\label{definition_of_H_n_t}
\mathcal{H}_n(v, t) =  \int_{B^n} \delta\left(v -\sum_{i=1}^{n}  k_f(\mathbf{X}_i)\right)  \prod_{i=1}^{n}   \phi(\mathbf{X}_i, t)  d\mathbf{X}_i,
\end{equation}
while for $\tilde{a}_n(t)$ we have
\begin{equation}
\label{Poisson_distribution_t}
\tilde{a}_n(t) = \frac{\alpha^n [Q(t)]^n}{n!}e^{-\alpha}.
\end{equation}
The factor $[Q(t)]^n$ appearing in the definition of $\tilde{a}_n(t)$ reflects the fact that we consider only those bubbles that contain excited fluorophores at time $t$. If we neglect spontaneous decay (\ref{reaction_3rd_quenching}), the probability that a fluorophore surrounded by $n$ independent quenchers is in the excited state at time $t$ is equal to $[Q(t)]^n$. From the equations (\ref{q_n_product_form}), (\ref{Poisson_distribution}), (\ref{definition_of_H_n}), (\ref{definition_of_H_with_H_n}), and (\ref{q_1_is_p_at_t_0}) - (\ref{Poisson_distribution_t}) it follows that
\begin{equation}
\label{H_t_at_zero_gives_H}
\mathcal{H}(v, 0) = {H}(v).
\end{equation}
For $t > 0$ the probability distribution $\tilde{a}_n(t)$ (\ref{Poisson_distribution_t}) is not normalized to unity, %
\begin{eqnarray}
\sum_{n=0}^{\infty} \tilde{a}_n(t)  &=& \exp\left[ c \int_{B} \left(S^{(e)}_{1}(t|\mathbf{X})-1 \right) g(\mathbf{X}) d\mathbf{X} \right] \nonumber \\  &=& \exp\left[ c \int_{B} \left[n(\mathbf{X}, t) - n(\mathbf{X}, 0) \right] d\mathbf{X} \right] = \tilde{I}(t), \nonumber \\ 
\label{sum_of_tilde_a}
\end{eqnarray}
and so we have
\begin{equation}
\label{lack_of_normalization_of_H_n_t}
\int_{0}^{\infty} \mathcal{H}(v, t) d v  = \tilde{I}(t) = e^{t/\tau_0} I(t)  \leq 1.
\end{equation}
The normalized counterpart of $\mathcal{H}(v, t)$ is therefore
\begin{equation}
\label{J_t_definition}
\mathcal{J}(v, t) = \frac{\mathcal{H}(v, t)}{\int_{0}^{\infty} \mathcal{H}(v, t) d v} = \frac{\mathcal{H}(v, t)}{\tilde{I}(t)}.
\end{equation}
$\mathcal{J}(v, t)$ (\ref{J_t_definition}) is a direct generalization of ${J}(v, t)$ (\ref{J_definition}). For example, one can check that Eq. (\ref{k_as_average_of_nu_wrt_J}) is still valid if $J(v, t)$ is replaced by $\mathcal{J}(v, t)$,
\begin{equation}
\label{k_as_average_of_nu_wrt_J_t}
c k(t) =   \int_{0}^{\infty} v \mathcal{J}(v, t) d v.
\end{equation}
Using (\ref{S_and_n_connection}), (\ref{definition_of_q_1_or_p_of_t}), (\ref{definition_of_phi_of_t}),  (\ref{q_1_is_p_at_t_0}), and (\ref{definition_of_H_n_t}), we can rewrite $\mathcal{H}(v, t)$ as given by (\ref{definition_of_H_t_with_H_n_t}) in the following form 
\begin{widetext}
\begin{eqnarray}
\label{definition_of_H_t_with_H_n_t_alternative_n_S}
\mathcal{H}(v, t) &=& e^{-\alpha} \sum_{n}  \frac{c^n}{n!} \int_{B^n} \delta\left(v -\sum_{i=1}^{n}  k_f(\mathbf{X}_i)\right)  \prod_{i=1}^{n}  n(\mathbf{X}_i, t)  d\mathbf{X}_i = e^{-\alpha} \sum_{n}  \frac{\alpha^n}{n!} \int_{B^n} \delta\left(v -\sum_{i=1}^{n}  k_f(\mathbf{X}_i)\right)  \prod_{i=1}^{n} S^{(e)}_{1}(t|\mathbf{X}_i) {q_1(\mathbf{X}_i)} d\mathbf{X}_i.  \nonumber \\
\end{eqnarray}
For $t=0$ we have $S^{(e)}_{1}(0|\mathbf{X})=1$. Recalling (\ref{definition_of_H_n}) and (\ref{definition_of_H_with_H_n}), it is then easy to see from (\ref{definition_of_H_t_with_H_n_t_alternative_n_S}) that (\ref{H_t_at_zero_gives_H}) is indeed fullfiled. 
Earlier we stated that if fluorophores and quenchers are immobile, we have $\mathcal{H}(v, t) = {H}(v) e^{-v t}$. To show this, note that in such a situation we also have $S^{(e)}_{1}(t|\mathbf{X}_i) = e^{- k_f(\mathbf{X}_i) t} $ (see Eq. (\ref{S_1_BGSE_solution_no_diffusion})), and (\ref{definition_of_H_t_with_H_n_t_alternative_n_S}) can be rewritten as
\begin{eqnarray}
\label{definition_of_H_t_with_H_n_t_alternative_n_S_SS}
\mathcal{H}(v, t) &=& e^{-\alpha} \sum_{n}  \frac{\alpha^n}{n!} \int_{B^n} \delta\left(v -\sum_{i=1}^{n}  k_f(\mathbf{X}_i)\right)  \prod_{i=1}^{n} e^{- k_f(\mathbf{X}_i) t}  {q_1(\mathbf{X}_i)} d\mathbf{X}_i  \nonumber \\ &=& e^{-\alpha} \sum_{n}  \frac{\alpha^n}{n!} \int_{B^n} \delta\left(v -\sum_{i=1}^{n}  k_f(\mathbf{X}_i)\right) e^{-v t}  \prod_{i=1}^{n}  {q_1(\mathbf{X}_i)} d\mathbf{X}_i = {H}(v) e^{-v t}.
\end{eqnarray}
\end{widetext}
In the above we have used the definition of $H(v)$ given by (\ref{definition_of_H_n}) and (\ref{definition_of_H_with_H_n}) as well as the following identity obeyed by Dirac's delta distribution: $h(x)\delta(x-x_0) = h(x_0)\delta(x-x_0)$, where $h(x)$ is an arbitrary function.

The crucial point now is that only for immobile fluorophores and quenchers we have $\mathcal{H}(v, t) = {H}(v)e^{-vt}$, and only then can ${H}(v)$ (and therefore $\mathcal{H}(v, t)$) be obtained from $\tilde{I}(t)$ by using Eq. (\ref{I_expressed_by_H}). However, if the fluorophores or quenchers are free to move, $\mathcal{H}(v, t)$ cannot be determined from the experimental data of the luminescence decay law $I(t)$. This is because in such a case Eq. (\ref{lack_of_normalization_of_H_n_t}) does not have the form of the Laplace transform with time $t$ playing the role of the Laplace variable. Therefore, one cannot use the inverse Laplace transform to compute $\mathcal{H}(v, t)$ from the knowledge of $I(t)$ - there is no generalization of Eq. (\ref{H_in_terms_of_I}).

On the one hand, the fact that in the presence of fluorophore or quencher diffusion neither $\mathcal{H}(v, t)$ nor $\mathcal{J}(v, t)$ can be determined from $I(t)$ is the obvious drawback of these distributions. On the other hand, we simply do not need them at all to predict values of experimentally observable quantities like $I(t)$ and $k(t)$ based on knowledge of the microscopic quantities describing the system. The construction of $\mathcal{H}(v, t)$ and $\mathcal{J}(v, t)$ is only an intermediate step on the way from microscopic quantities to macroscopic observables, a step that can be skipped altogether. Nevertheless, the time-dependent rate constant distributions $\mathcal{H}(v, t)$ and $\mathcal{J}(v, t)$ are worth discussing, since an analysis of their construction and properties sheds additional light on the rate distribution formalism. Moreover, time-dependent rate constant distributions have been used in the literature. \cite{kuzmin2008evolution}

\subsubsection{Time-averaged rate distribution $\overline{\mathcal{H}(v)}$}

We know that $\mathcal{H}(v, 0) = H(v)$, see (\ref{H_t_at_zero_gives_H}).  Is it possible to get $H(v)$ from $\mathcal{H}(v, t)$ in an alternative way -- by some kind of time averaging? Within the original formulation of the rate distribution formalism we have the following identity
\begin{equation}
\label{time_averaged_H_exp_definition}
\int_{0}^{\infty} H(v) e^{-v t} dt = \frac{H(v)}{v}.
\end{equation}
Now in (\ref{time_averaged_H_exp_definition}) we replace ${H}(v) e^{-v t}$ by its generalization: $\mathcal{H}(v, t)$ and define
\begin{equation}
\label{time_averaged_H_t_definition}
\int_{0}^{\infty} \mathcal{H}(v, t) dt = \overline{\mathcal{H}(v)}.
\end{equation}
We have  
\begin{equation}
\label{time_averaged_H_t_is_not_normalized}
\int_{0}^{\infty}  \overline{\mathcal{H}(v)} dv = \int_{0}^{\infty} \tilde{I}(t) dt.
\end{equation}
In general, $\int_{0}^{\infty} \tilde{I}(t) dt \neq 1$, so $\overline{\mathcal{H}(v)}$ is not properly normalized. However, it is easy to check that the following function
\begin{equation}
\label{time_averaged_H_t_times_v_is_normalized_chi}
\chi(v) \equiv v   \overline{\mathcal{H}(v)} 
\end{equation}
is properly normalized: $\int_{0}^{\infty}  \chi(v) dv =1$ and $\chi(v)$ can therefore be regarded as a probability distribution. One can now ask if it is  possible that
\begin{equation}
\label{time_averaged_H_t_giving_H}
H(v)  = \chi(v)?
\end{equation}
As one might guess, the answer is affirmative only if the fluorophores and quenchers are immobile. To show this, it is convenient to first compute the following Laplace transform of $\mathcal{H}(v, t)$, 
\begin{widetext}
\begin{eqnarray}
\label{definition_of_G_t_with_G_n_t_alternative_n_S}
\mathcal{L}[ H(v)](t) &=& \int_{0}^{\infty}  \mathcal{H}(v, t) e^{-s v } d v = e^{-\alpha} \sum_{n}  \frac{\alpha^n}{n!} \int_{0}^{\infty}  \int_{B^n} \delta\left(v -\sum_{i=1}^{n}  k_f(\mathbf{X}_i)\right) e^{-s v} \prod_{i=1}^{n} S^{(e)}_{1}(t|\mathbf{X}_i) {q_1(\mathbf{X}_i)} d\mathbf{X}_i dv \nonumber \\ &=& e^{-\alpha} \sum_{n}  \frac{\alpha^n}{n!} \int_{0}^{\infty}  \int_{B^n} e^{- s \sum_{i=1}^{n}  k_f(\mathbf{X}_i)} \prod_{i=1}^{n} S^{(e)}_{1}(t|\mathbf{X}_i) {q_1(\mathbf{X}_i)} d\mathbf{X}_i   = \exp\left \{c \int_{B} \left[e^{- s  k_f(\mathbf{X})} S^{(e)}_{1}(t|\mathbf{X})-1 \right] g(\mathbf{X})  d\mathbf{X} \right\} \nonumber \\  &=& \exp\left\{c \int_{B} \left[e^{- s  k_f(\mathbf{X})} n(\mathbf{X}, t) - n(\mathbf{X}, 0) \right]  d\mathbf{X} \right\} \equiv \mathcal{G}(s, t) \equiv \exp \left[ \mathcal{Z}(s, t) \right],
\end{eqnarray}
\end{widetext}
which, except for the sign of $s$, is the time-dependent moment generating function for $\mathcal{H}(v, t)$. Note that now the Laplace variable is no longer the observation time $t$, and $\mathcal{G}(s, t)$ has no physical interpretation, it is just a useful mathematical quantity. We have 
\begin{equation}
\mathcal{G}(0, t) = \tilde{I}(t),
\end{equation}
so the fact that $\mathcal{H}(v, t)$ is not properly normalized (see equation (\ref{lack_of_normalization_of_H_n_t})) is correctly encoded in $\mathcal{G}(s, t)$ (\ref{definition_of_G_t_with_G_n_t_alternative_n_S}).

Coming back to (\ref{time_averaged_H_t_giving_H}), we want to know when the  equality holds. Using generating functions, the condition (\ref{time_averaged_H_t_giving_H}) can  be expressed in the following equivalent form 
\begin{eqnarray}
\label{time_averaged_H_t_giving_H_with_G}
\tilde{I}(s) &=& \int_{0}^{\infty} H(v) e^{-s v } d v = \int_{0}^{\infty}   \chi(v) e^{-s v } d v  \nonumber \\  &=&  \int_{0}^{\infty}  \int_{0}^{\infty} v \mathcal{H}(v, t) e^{-s v } dt d v = -\frac{\partial}{\partial s} \int_{0}^{\infty} \mathcal{G}(s, t) dt.  \nonumber \\
\end{eqnarray}
In (\ref{time_averaged_H_t_giving_H_with_G}) we have used (\ref{time_averaged_H_t_times_v_is_normalized_chi}), the basic properties of the Laplace transforms, and the fact that the order of integration with respect to $s$ and $t$ can be interchanged. If the positions of the fluorophores and quenchers are fixed, we have $e^{- t k_f(\mathbf{X})}n(\mathbf{X}, 0) = n(\mathbf{X}, t)$. In this case $\mathcal{Z}(s, t)$ defined by (\ref{definition_of_G_t_with_G_n_t_alternative_n_S}) simplifies to
\begin{equation}
\mathcal{Z}(s, t) = c \int_{B} \left[e^{- s k_f(\mathbf{X})} e^{- t k_f(\mathbf{X})} - 1 \right] n(\mathbf{X}, 0) d\mathbf{X}
\end{equation}
so it is a symmetric function of its variables: $\mathcal{Z}(s, t) = \mathcal{Z}(t, s)$. Therefore, $\mathcal{G}(s, t)$ is also symmetric. Moreover, a much stricter condition is satisfied -- the partial derivatives of $\mathcal{Z}(s, t)$ with respect to $s$ (denoted $\mathcal{Z}^{\prime}(s, t)$) and with respect to $t$ (denoted $\dot{\mathcal{Z}}(s, t)$) are also symmetric and equal to each other:
\begin{eqnarray}
\label{symmetry_of_Z}
\mathcal{Z}^{\prime}(s, t) = \mathcal{Z}^{\prime}(t, s) & = &  \dot{\mathcal{Z}}(s, t) = \dot{\mathcal{Z}}(t, s).
\end{eqnarray}
The same conditions are satisfied by partial derivatives of $\mathcal{G}(s, t)$. Now using (\ref{symmetry_of_Z}) we can easily prove (\ref{time_averaged_H_t_giving_H_with_G}). We have
\begin{eqnarray}
\label{time_averaged_H_t_giving_H_with_G_bis}
-\frac{\partial}{\partial s} \int_{0}^{\infty} \mathcal{G}(s, t) dt & = & - \int_{0}^{\infty} \mathcal{G}^{\prime}(s, t) dt   \nonumber \\ = - \int_{0}^{\infty} \dot{\mathcal{G}}(s, t) dt & = & \mathcal{G}(s, 0) = \mathcal{G}(0, s)  = \tilde{I}(s),
\end{eqnarray}
hence the condition (\ref{time_averaged_H_t_giving_H_with_G}) is satisfied. But if fluorophores or quenchers are free to move, it is no longer generally true that $e^{- t k_f(\mathbf{X})}n(\mathbf{X}, 0) = n(\mathbf{X}, t)$, neither $\mathcal{G}(s, t)$ nor $\mathcal{Z}(s, t)$ is a symmetric function with symmetric and equal partial derivatives, and the above derivation breaks down. Therefore, in such a case, (\ref{time_averaged_H_t_giving_H}) is also generally not true.

\section{Conclussions}

Within the standard variant of the rate-distribution formalism, both the decay rates themselves and their probability distributions have a well-defined physical interpretation only when fluorophores and quenchers are immobile. In this case, the probability distributions of the rate constants can be calculated from knowledge of the initial ($t=0$) microscopic properties of the system, such as the distribution of quenchers around the fluorophore and the microscopic transfer rates.

However, if fluorophores or quenchers are allowed to move (either by diffusion or by some deterministic motion), we lose the physical interpretation of the rate constants and the connection between microscopic details of the system and observable macroscopic quantities such as luminescence. 

Therefore, time-independent continuous or discrete distributions of rate constants should not be used to describe luminescence decay not only in simple liquids, but also in many soft matter systems. This is because in such systems the relative motion of fluorophore and quencher molecules usually cannot be neglected. The same remark applies to the decay time distributions, since there is a one-to-one correspondence between each such distribution and the corresponding distributions of the rate constants. 

The connection between microsopic and macroscopic quantities, as well as the sound physical interpretation of the rate constants, can be restored by introducing the time-dependent rate constant distributions. In such a case, however, these distributions cannot be obtained from the experimentally measured luminescence decay and are therefore of limited practical use. To construct the time-dependent rate constant distributions that are correct even in the case of fluorophore or quencher motion, we must rely on mathematical models such as the backward Smoluchowski equation or differential encounter theory.

This conclusion can be generalized to any situation where the survival probability of the excited state is not exponential, even if the phenomenon is unimolecular.

\section*{Acknowledgments}

The National Science Centre OPUS Grant No. 2019/33/B/ST4/01443 made this work possible.  

\appendix

\section{Appendix: General scenario leading to the rate distribution formalism \label{Appendix_General_Scheme}}

When fluorophores and quenchers are immobile, the rate distribution formalism can be derived in a way that is heuristic, formal, but very general - it does not depend on the microscopic details of a particular system. More specifically, in this appendix we show that the rate distribution formalism (RDF) of Berberan-Santos and co-workers can be obtained as a continuous limit of the simple kinetic model. The arguments presented below also show the "mean-field", averaged nature of the probabilities of the RDF. Therefore, this approach should be used with caution when the number of fluorophore molecules in the system of interest is small (the precise meaning of "small" depends on the system under consideration).

Consider a molecule (or an atom) that can be described by a \textit{finite} number $\mathcal{S} + 1$ of discrete states $\Xi_{\ell}$, $\ell = 0, 1, 2, \ldots, \mathcal{S}$. Let us call $\Xi_{0}$ the ground state and states with $\ell = 1, 2, \ldots, s$ the excited states. These may or may not be actual energy levels of the molecule itself. $\Xi_{\ell}$ can also refer to the states of our molecule taken together with its 'microenvironment', i.e. the molecules surrounding it. In any case, we assume that only the following first-order transitions are possible
\begin{eqnarray}
\label{reaction_appendix}
\Xi_{\ell}   & \xrightarrow{v_{\ell}}  &  \Xi_{0},
\end{eqnarray}
where $\ell = 1, 2, \ldots, \mathcal{S}$ and $v_{\ell}$ are the corresponding rates.

Of particular importance to us is the situation where the molecule of interest is the excited fluorophore surrounded by a number of quenchers. The crucial assumption we make is that both fluorophores and quenchers are immobile -- their positions do not change with time. Then each $\Xi_{\ell}$ with $\ell > 0$ corresponds to a subset of the excited fluorophore microenvironments for which $v_{\ell} = \sum_{i=1}^n k_f(\mathbf{X}_i)$, and each such microenvironment is defined by the number $n$ and the positions $\mathbf{X}^n$ of the quenchers. (These variables determine the value of the transfer rates $k_f(\mathbf{X}_i)$ and the rate constant $v=v_{\ell}$ via the equation (\ref{nu_definition_n_bubble})).

Now let $\mu_{\ell}(t)$ be the number of excited fluorophore molecules in the $\ell$-th state at time $t$, $\ell \geq 1$. The total number of excited molecules in our system is equal to 
\begin{equation}
\label{N_as_a_sum_of_mu_l}
N(t) = \sum_{\ell=1}^{\mathcal{S}} \mu_{\ell}(t).
\end{equation}
We are only interested in the excited molecules and ignore those that remained in the ground state after the pulse excitation applied at $t=0$, hence $\mu_{0}(0)=0$.  Both $\mu_{\ell}(t)$ and $N(t)$ change with time as the molecules transition to the ground state ($\ell = 0$) either by quenching reaction (\ref{reaction_3rd_quenching}) or by spontaneous emission (\ref{reaction_2nd_spontaneous_emission}).

First, for simplicity, we assume that for all values of $\ell = 1, 2, \ldots \mathcal{S}$ the populations (occupation numbers) $\mu_{\ell}(t)$ are large enough that we can treat them as continuous variables (concentrations). In other words, we assume that  
\begin{equation}
\label{condition_of_continuity_for_mu_l}
\forall_{1 \leq \ell \leq \mathcal{S}}: ~~~ \mu_{\ell}(t) \gg 1.
\end{equation}
In such a case, according to Eq. (\ref{reaction_appendix}), the time evolution of each $\mu_{\ell}(t)$ can be described by the following first-order deterministic kinetic equation 
\begin{equation}
\label{mu_l_ODE_deterministic_kinetics}
\frac{d \mu_{\ell}(t)}{dt} = - \tilde{\nu}_{\ell}  \mu_{\ell}(t) = - \left( \nu_{\ell} + \frac{1}{\tau_0} \right) \mu_{\ell}(t).
\end{equation}
Therefore we get
\begin{equation}
\label{mu_l_t}
\mu_{\ell}(t) = \mu_{\ell}(0) e^{-  t/\tau_0} e^{- \nu_{\ell} t}.
\end{equation}
A more formal and rigorous justification for the use of continuous $\mu_{\ell}(t)$ variables is provided by the formalism of the Chemical Master Equation (CME), which allows each $\mu_{\ell}(t)$ to be replaced by the average value of the corresponding discrete molecule number, see below. Now let us define 
\begin{equation}
\label{GS_of_H_k_definition_of_h_l}
h_{\ell} =  \frac{\mu_{\ell}(0)}{\sum_{\ell=1}^{\mathcal{S}} \mu_{\ell}(0)} = \frac{\mu_{\ell}(0)}{N(0)}.
\end{equation}
Obviously, $\sum_{\ell=1}^{\mathcal{S}} h_{\ell} = 1$ and $\forall_{1 \leq \ell \leq \mathcal{S}}: h_{\ell} \geq 0$, so $h_{\ell}$ can be interpreted as a probability distribution; more precisely, as the discrete counterpart of $H(\nu)$ (\ref{H_in_terms_of_I}). From (\ref{N_as_a_sum_of_mu_l}), (\ref{mu_l_t}) and (\ref{GS_of_H_k_definition_of_h_l}) we also have 
\begin{equation}
\label{I_expressed_by_h_l}
I(t) \equiv \frac{N(t)}{N(0)} = \frac{\sum_{\ell=1}^{\mathcal{S}} \mu_{\ell}(t)}{\sum_{\ell=1}^{\mathcal{S}} \mu_{\ell}(0)} =   e^{-  t/\tau_0} \sum_{\ell=1}^{\mathcal{S}} h_{\ell}e^{- \nu_{\ell} t}.
\end{equation}
In the limit of large $\mathcal{S}$ we can replace the sum by an integral: 
\begin{equation}
\label{}
\sum_{\ell=1}^{\mathcal{S}} h_{\ell}e^{- \nu_{\ell} t} \longrightarrow \int_{0}^{\infty} H(\nu) e^{-\nu t} d \nu,
\end{equation}
and (\ref{I_expressed_by_h_l}) gives the equation (\ref{I_expressed_by_H}). The discrete counterpart of the time-dependent distribution $J(\nu, t)$ (\ref{J_definition}) also has a very simple and natural definition:
\begin{equation}
\label{GS_of_H_k_definition_of_j_l}
j_{\ell} \equiv \frac{\mu_{\ell}(t)}{N(t)} = \frac{h_{\ell}e^{- \nu_{\ell} t}}{\sum_{\ell=1}^{\mathcal{S}} h_{\ell}e^{- \nu_{\ell} t}},
\end{equation}
which actually coincides with (\ref{J_definition}) in the limit of large $\mathcal{S}$.

So far we have assumed that all state populations $\mu_{\ell}(t)$ (which are in fact natural numbers) are so large that they can be approximated by continuous variables whose time evolution is governed by Eq. (\ref{mu_l_ODE_deterministic_kinetics}). But clearly this is not necessarily the case, especially for large values of $\mathcal{S}$ (large number of states), where the population of a particular state $\ell \geq 1$ cannot be too large due to the constraint imposed by Eq. (\ref{N_as_a_sum_of_mu_l}). Also at the $t \to \infty$ limit, values of at least some $\mu_{\ell}(t)$ variables may be close to zero. However, in such a case, the above reasoning and all its derivations can be made perfectly valid by invoking the Chemical Master Equation (CME) formalism. Within this approach, instead of the deterministic kinetic equation (\ref{mu_l_ODE_deterministic_kinetics}), we use the following CME \cite{mcquarrie1967stochastic}
\begin{equation}
\label{mu_l_CME_stochastic_kinetics} 
\frac{dP_{\ell}(m_{\ell}, t)}{dt} = \tilde{\nu}_{\ell} (m_{\ell} + 1)P_{\ell}(m_{\ell} + 1, t) - \tilde{\nu}_{\ell} m_{\ell} P_{\ell}(m_{\ell}, t)
\end{equation}
with the deterministic initial condition
\begin{equation}
\label{CME_initial_condition}
P_{\ell}(m_{\ell}, t) = \begin{cases}
  1  & \text{ if} ~~~ m_{\ell} = M_{\ell}, \\
  0 & \text{ if} ~~~ m_{\ell} < M_{\ell}.
\end{cases}
\end{equation}
Again $\tilde{\nu}_{\ell} = {\nu}_{\ell} + 1/\tau_0$, while $m_{\ell} = 0, 1, \dots, M_{\ell}$ is the number of fluorophore molecules in the state $\ell$. $P_{\ell}(m_{\ell}, t)$ is the probability that at time $t$ we have exactly $m_{\ell}$ fluorophore molecules in the state $\ell$, provided that at $t=0$ we had exactly $M_{\ell}$ such molecules as given by (\ref{CME_initial_condition}). We also have $P_{\ell}(M_{\ell}+1, t) = 0$. The equation (\ref{mu_l_CME_stochastic_kinetics}) can be solved,  \cite{mcquarrie1967stochastic} one gets
\begin{equation}
\label{mu_l_CME_stochastic_kinetics_solution}
P_{\ell}(m_{\ell}, t) ={M_{\ell} \choose m_{\ell}}\left(1-e^{-\tilde{\nu}_{\ell} t}\right)^{M_{\ell} - m_{\ell}} \left( e^{-\tilde{\nu}_{\ell} t}\right)^{m_{\ell}}.
\end{equation}
But what we need now is just the fact that for the first order reaction described by (\ref{mu_l_CME_stochastic_kinetics}), the average number of molecules, i.e.
\begin{equation}
\label{definition_of_average_m}
\mathcal{A}_{\ell}(t)  \equiv \sum_{m_{\ell} =0}^{M_{\ell}} m_{\ell} P_{\ell}(m_{\ell}, t)
\end{equation}
obeys the same time evolution equation (\ref{mu_l_ODE_deterministic_kinetics}) as the continuous concentration $\mu_{\ell}(t)$ of the deterministic approach:
\begin{equation}
\label{average_m_l_ODE_deterministic_kinetics}
\frac{d \mathcal{A}_{\ell}(t)}{dt} = - \tilde{\nu}_{\ell} \mathcal{A}_{\ell}(t),
\end{equation}
and therefore 
\begin{equation}
\label{average_m_l_ODE_deterministic_kinetics_solution}
 \mathcal{A}_{\ell}(t) = M_{\ell} e^{ - \tilde{\nu}_{\ell} t } =  M_{\ell} e^{-  t/\tau_0} e^{- \nu_{\ell} t},
\end{equation}
$\mathcal{A}_{\ell}(0) = M_{\ell}$. To make the derivation of the distributions $h_{\ell}$ (\ref{GS_of_H_k_definition_of_h_l}) and $j_{\ell}$ (\ref{GS_of_H_k_definition_of_j_l}) and the formula (\ref{I_expressed_by_H}) completely rigorous, it is sufficient to use $\mathcal{A}_{\ell}(t)$ (\ref{definition_of_average_m}) instead of $\mu_{\ell}(t)$ in both (\ref{GS_of_H_k_definition_of_h_l}) and (\ref{GS_of_H_k_definition_of_j_l}).

\bibliography{bibliography_RDF_vs_DRM}

\end{document}